%Paper: hep-th/9410011
%From: bershad@string.harvard.edu (Michael Bershadsky)
%Date: Mon, 3 Oct 1994 15:16:35 -0400

\input harvmac

\def\boxit#1{\vbox{\hrule\hbox{\vrule  \kern3pt
\vbox{\kern-12pt #1\kern-10pt}\kern3pt\vrule}\hrule}}

\def\Boxit#1{\vbox{\hrule\hbox{\vrule  \kern3pt
\vbox{\kern-15pt #1\kern-7pt}\kern3pt\vrule}\hrule}}

\Title{\vbox{\hbox{HUTP--94/A013}} }
{\vbox{\centerline{Theory of K\"ahler Gravity}  }}

\centerline{M. Bershadsky$^{\,1}$ and V. Sadov$^{\,1,2}$}
\vskip .3in
 \centerline{$^{1\,}$Lyman Laboratory of Physics, Harvard
University}
\centerline{Cambridge, MA 02138, USA}
\bigskip
\centerline{$^{2\,}$L.~D.~Landau Institute for Theoretical Physics}
\centerline{Russia}
\vskip .3in

In this paper we discuss the  connection on a space of $N=2$ TCFT's
that
appears in the context of  background (in)dependence.
We  formulate a family of  {\it target space field theories} with a
similar
connection on it. Each theory  is a gauge theory (with the gauge
group being
${\cal SD}iff $ in the case of $3$-fold). It describes deformations
of K\"ahler
structures much like Kodaira Spencer theory describes deformations of
the
complex structures.
It is manifestly background independent. It appears to be a target
space field
theory for supersymmetric quantum mechanics.

%\draft
\Date{}
\catcode`\|=12

\newsec{Introduction}

Kodaira-Spencer theory  \ref\bcov{
M. Bershadsky, S. Cecotti, H. Ooguri and C. Vafa, Kodaira-Spencer
theory of
Gravity and Exact results for Quantum Strings Amplitudes, to appear
in CMP}
is a string field theory for topological $B$-model.
As it was noticed in \ref\wit{E. Witten, in {\it Essays
on Mirror
Manifolds} ed. by S. T. Yau,  International Press, 1992} in this case
the {\it
string field theory} reduces to a
{\it field theory}. The reason for this is that topological $B$-model
coupled to gravity is essentially independent of the
K\"ahler structure. Rescaling the volume to infinity one recovers
that  the
path integral is dominated by
highly degenerate Riemann surfaces. One can think of  degenerate
Riemann
surfaces
as infinitely thin tubes attached to each other. In other words,
topological
$B$-model
can be described as supersymmetric quantum mechanics.
In the case of topological $A$-model the situation is different.
It is known that nontrivial worldsheet configurations  (instantons)
play the
crucial role
in topological $A$-model. String field theory for $A$-model is
defined on the
loop space.
In the large volume limit the instanton effects are suppressed and
one can
describe
the semiclassical limit of string field theory as supersymmetric
quantum
mechanic (SQM).
This SQM makes sense by itself even when the volume is not large.
It also exhibits some properties of underlying string theory.

SQM in question describes deformations of K\"ahler structures in the
same
way as
Kodaira-Spencer theory \bcov\
describes deformations of complex structures. We will call this
theory AKS,
where A stands for topological
A-model in Witten's terminology  \wit,
 and KS stands for
K\"ahler
structures. It is known that  the perturbation
theory of Chern-Simons theory can
be interpreted as a perturbation theory of open strings propagating
on $T^*(M)$, where $M$ is three dimensional\ref\witcs {E. Witten,
{\it Chern--Simons Gauge Theory as a String Theory},
IASSNS-HEP-92/45,
hep-th/9207094}.
In trying to describe the closed string sector (which is required by
consistency
in open string theory) E. Witten introduced the action for AKS theory
\witcs.
In spite of the fact that AKS is very similar to Chern-Simons it is
not a
topological theory.
Its Hamiltonian is non trivial, while the phase space is  finite
dimensional.
On the other hand the AKS theory enjoys the properties of being
independent on complex structure. It depends only on the K\"ahler
{\it class}
of the
metric. We call this theory a K\"ahler topological theory defined on
a K\"ahler
manifold.
The gauge invariant observables of Chern-Simons theory are Wilson
lines.
In \witcs\ the Wilson lines were used in order to  incorporate
the worldsheet instantons in string theory.
In the case of AKS theory
we do not know any gauge invariant
observables\foot{This situation is very similar to
the conventional theory of gravity} except  the action.
It is tempting to suggest  that the would be gauge invariant
observables
are related to holomorphic curves in the target space, or saying
differently to worldsheet instantons.

The plan of this paper is the following.
In  Section $2$ we discuss the notion of background independence.
This discussion is quite general and is applicable to
any $N=2$ topological conformal field theories
(TCFT).
There is a natural connection on the moduli space of  TCFT.
This connection allows one to identify the {\it perturbed} TCFT at
certain
background
with {\it unperturbed} TCFT at another background.
Background independence is equivalent to the statement that
the connection is flat.  In general, there is an obstacle known as
holomorphic
anomaly.
In order to avoid this problem one has to consider only the
{\it holomorphic} deformations of TCFT.
The  background independence imposes strong restrictions on the form
of contact
terms.
In principle, these equations should fix the connection in full
(quantum)
theory.
Semiclassically, these equations have a unique solution and
supersymmetric
quantum mechanics
(SQM) is a theory which solves them.

Whereas the moduli space of TCFT is a complexified
K\"ahler cone,  the moduli space of  SQM is a real
K\"ahler cone.  As explained in Section 2 we
identify  the   {\it real}  deformations of  SQM
with the {\it holomorphic} deformations of $N=2$ TCFT by means of
analytic continuation. Under this identification the semiclassical
limit of thr holomorphic connection of TCFT is mapped on the
flat connection of SQM. Therefore the holomorphic anomaly
does not show up in SQM.  As a result the SQM is background
independent.
This connection has a natural geometric interpretation.

Sections $3-5$ are devoted to AKS theory and its properties.
One can construct AKS action
for a given point in the
moduli space of K\"ahler structures and the tangent vector ($\omega$
and $x \in
H^2$) that serves as the
background data.
AKS is a  gauge invariant theory with symmetries generated by the
large
volume limit of string BRST $Q$. The classical equation of motion is
equivalent to the condition $Q^2=0$. The solution of this equation of
motion
determines a perturbed K\"ahler structure (the precise meaning of
this will
be explained). The gauge group is non-abelian and in the case of
$3$-dimensional
K\"ahler manifolds is isomorphic to volume
preserving diffeomorphisms.
In Section 3  we discuss the Batalin-Vilkovisky formalism for AKS
theory.
The absence of higher Massey products on
the K\"ahler manifold makes AKS action exact at the quantum level.
In Section $5$  we discuss the Hamiltonian quantization of AKS
theory.

AKS is  a target space field theory for suitably modified
(along the lines of  reference \ref\yank{N. Markus and S.
Yankielowicz,
The topological B model as a twisted spinning particle,
CERN-TH-7402/94,
TUP-2192-94, hep-th/9408116})
$N=2$ SQM.
The connection discussed above in the context of SQM naturally
appears in AKS. It allows to relate AKS theories at different
K\"ahler  structures. The idea of background independence can
be fully applied to AKS.  Under the variation of K\"ahler structure
the AKS action minus the action evaluated on the classical trajectory
scales with volume as the second power.
This scaling can be reabsorbed into the
redefinition of the coupling constant $g^2 \rightarrow g^2 \cdot {\rm
Vol}^2$.
This power {\it differs} from the one naively expected from the
dimensional
considerations.

We conjecture that the  effective action $\Gamma(x)$ for SQM is a
free energy
for
AKS which depends on $x$ as a parameter.
In Section 6 we prove this relation at the tree level.
In doing this we found a very simple mechanism that allows one to
rewrite
 {\it vacuum} diagrams for AKS as S-matrix diagrams of effective
field theory.

\newsec {Contact terms}
\subsec{Background independence and the contact term algebra}

Consider  a family of  A-twisted $N=2$ superconformal $\sigma$-models
on
Calabi-Yau space.
Let us start with the discussion  of how the susy generators vary
under the
variation along the moduli. The physical operators ({\it chiral
fields}) in
topological A-model are given by
BRST cohomology
\eqn\physop{\phi=\phi_{i_1 ...i_n \bar j_1...\bar j_m} \chi^{i_1} ...
\chi^{i_n} \overline \chi^{j_1} ...
              \overline \chi^{j_m}.}
and are independent of the target space metric. These are 0-forms on
the world
sheet.  Solving the descent equation \ref\Sing{I.~Singer, L.~Balieu,
Annecy Field Theory (1988) 12;
Commun. Math. Phys. 125 (1989) 227}  one
obtains the operators $\phi ^{(1)}$ and $\phi ^{(2)}$, respectively
1- and
2-forms on the world sheet which are BRST-closed only modulo total
derivative.
Also,  we will need to  consider the {\it antichiral fields}
\eqn\anch{\phi ^\dagger=\phi_{i_1 ...i_n \bar j_1...\bar j_m}
\rho^{i_1} ...
\rho^{i_n} \overline \rho^{j_1} ...
              \overline \rho^{j_m}.}
The operator \anch\ is a $(n,m)$-form on the world sheet. The
antichiral fields
are not BRST-closed.  In this paper we will restrict ourselves to
studying the
effects  of deforming the theory by  exactly marginal  operators
\physop\
(~corresponding to the target space $(1,1)$-forms~) and by their
antichiral
counterparts \anch:
\eqn\action{S \rightarrow S + \epsilon^a \int \phi^{(2)}_a+\bar
\epsilon^a
\int [Q,[\bar Q,\phi^{\dagger} _a ]].}
In \action\ the perturbation $\int \phi^{(2)}_a$ is exactly marginal,
while
$\int [Q,[\bar Q,\phi^{\dagger} _a ]]$  is BRST trivial.

For a given point $p$ in the moduli space\foot{Talking about the
moduli space
here, we usually mean the complexified cone of  K\"ahler classes
${\rm K}_{\bf
C}$.} ${\cal M}$  the perturbations are the vectors in the tangent
space ${\cal
T}_p{\cal M}$ to $\cal M$ at $p$. The coordinates $\epsilon ^a,\,
\bar \epsilon
^a$ make  ${\cal T}_p{\cal M}$ into a complex space.  At the moment,
we have a
family of {\it perturbed}  topological theories \action,
parameterized
by points
$(p,\,\epsilon ,\,\bar \epsilon )$ of the tangent bundle ${\cal TM}$
to the
moduli space ${\cal M}$.
The concept of  {\it global} background
independence is that  any perturbed theory  at
$(p,\,\epsilon ,\,\bar \epsilon ) \in {\cal TM}$ is equivalent to
some
unperturbed theory  at
$(\tilde p,\,0 ,\,0 )$, where the coordinates of
$\tilde p \in {\cal M}$
are functions of $(\epsilon ,\,  \bar \epsilon)$ and
$p$.
In fact there is a whole family of   perturbed
theories $(p', \epsilon (p'), \bar \epsilon (p') )$ parameterized by
$p'$
such that
$\epsilon(p)=\epsilon$, $\bar \epsilon(p)=\bar \epsilon$ and
$\epsilon(\tilde p)=\bar \epsilon(\tilde p)=0$.
This implies  existence of  a connection
${\cal D}$ on
${\cal TM}$ such that
\item{$\bullet$} It preserves the physical content of  theory. This
means that
the parallel transport does not alter  the correlation functions of
the theory;
\item{$\bullet$} Every constant section
$x=(\epsilon (p), \bar \epsilon(p))$
(a solution of the equation ${\cal D} x=0$)
has one and only one zero on  ${\cal M}$.

\noindent
The first  condition above guarantees that  the correlators remain
the same on
the constant section passing through a given perturbed theory. The
second
condition allows one  to reach the
unperturbed theory
unambiguously moving along the constant section.
The connection should be necessarily
{\it flat}. The very notion of constant section having a  zero at the
particular point requires this.

In general there is {\it no} flat connection with all the above
properties.
There is a non-zero curvature, which is given by the relation of
special
geometry.
On the other hand the tangent bundle is holomorphic and  $(0,2)$ and
$(2,0)$
components  of the curvature
are equal to zero which means that holomorphic (antiholomorphic)
directions are flat.
Therefore, perturbing the theory only by chiral
primary fields
one can consistently define the path-independent parallel transport
of the
tangent space.
This perturbation by chiral
primary fields is nothing else but an analytic continuation in
holomorphic
direction.

In general, such connection is {\it affine} --- the transformation
it induces
in the tangent space is not linear but rather a composition of  the
linear one
and a translation. The linear piece provides a  {\it linear}
connection $D$ on
${\cal TM}$.

Leaving the discussion of  the global background independence for the
next
sections, here we will concentrate on the local problem.  By the {\it
local}
background independence it is  usually meant that it is really
possible to
identify  the deformations \action\ as tangent vectors to $\cal M$.
So from now
till the end of this section we assume that  the parameters
$\epsilon,\,
\bar \epsilon$ are infinitesimally small.

It is convenient to define the {\it Hilbert space bundle} ${\cal HM}$
as a
bundle over the
moduli space whose fiber at every point is given by Hilbert space.
The space of physical states with charges $(q, \bar q)=(1, \bar 1)$
can be
identified with the tangent space
to the moduli space and therefore ${\cal TM}$ is a subbundle in
${\cal HM}$.
As we will see below there are two connections --  $D_{\cal H}$ on
${\cal HM}$
and
$D$ on ${\cal TM}$. The connection $D$ can be obtained
by $\it restricting$ $D_{\cal H}$ on the tangent bundle.

Let us recall the basics of  the state-operator correspondence for
families of
topological A-models.
As mentioned above, the  operators are independent of the parameters
$\epsilon,\, \bar \epsilon$ of deformation.
Now, the state-operator correspondence implies that  the states do
depend on
$\epsilon,\, \bar \epsilon$.
Indeed,  the
state $|\psi \rangle$ (associated with the wave function $\psi$) is
given by
the path integral over hemisphere
$\Sigma$
with appropriate boundary conditions. Under the variation \action\ of
the
action
this path integral  varies according to
\eqn\stateop{ \delta|  \psi \rangle= \int_{\Sigma} \phi^{(2)}
d^2z~| \psi
\rangle }
In the case of exactly marginal deformations of conformal field
theory this
integral picks
just a contact term contribution  since the bulk term is zero
\eqn\contact{\phi^{(2)} (z) \psi (x) \sim \mu (\phi,\psi)\delta^{2}
(z-x) .}  Thus we obtain the equation
$ \delta  |  \psi \rangle = |\mu (\phi,\psi )\rangle ~$
describing deformations of the states by \action.
As will be clear below, the contact term $\mu (\phi ,\psi)$ is a
chiral
operator, {\it not} $Q$ closed in general even if $\psi$ is $Q$
closed.
This contact term defines a connection on the Hilbert space bundle
${\cal HM}$.

Before discussing the connections $D_{\cal H}$ and $D$ let us first
discuss the
variation of
 susy generators. The
relevant OPEs are
\eqna\ope{
$$
\eqalignno{
G^+ _z (z) \phi^{(2)}(x)&=\partial_x \Bigl({1 \over
x-z}\phi^{(0,1)}\Bigr)+
\delta ^2(x-z)\delta_\phi G^+ _z,   &\ope a \cr
G^+ _z (z) \phi^{\dagger (2)}(x)&= \delta ^2(x-z)\delta_\phi G^+ _z,
&\ope b \cr
G^- _{zz} (z) \phi^{\dagger(2)}(x)&=\partial_x \Bigl({1 \over
x-z}[\bar
Q,\phi^{\dagger}]\Bigr)+
\delta ^2(x-z)\delta_{\phi}G^- _{zz}, &\ope c \cr
G^- _{zz} (z) \phi^{ (2)}(x)&= \delta ^2(x-z)\delta_\phi G^- _{zz},
&\ope d \cr
}$$}
The contact terms in \ope a - \ope d\  (the coefficients in front of
the
$\delta$-functions) ensure the conservation of the perturbed
currents.
To interpret these OPEs we note first that the $N=2$ susy generators
$G^+ _z$
and $G^- _{zz}$
explicitly depend on the target space metrics. These variations
explicitly
appear
as coefficients in front of the $\delta$-functions.

To understand the importance of  the total derivatives in \ope a -
\ope d\  let
us
consider a BRST operator $Q$ acting on some state $| \psi \rangle$ in
the
$perturbed$ theory. Its action is given by
$\oint dz G^+ _z$, where integration runs over the boundary of the
hemisphere.
The perturbation  does not commute
with the BRST operator. It makes a difference whether
we first make a  perturbation and then compute the action of  BRST or
vice
versa.  The difference is given by
\eqn\change{ Q  |\delta _\phi \psi  \rangle =\epsilon  \oint
_{\partial \Sigma}
dz  G^+ _z   \int_{\Sigma} d^2x \phi^{(2)} |\psi \rangle =
 \epsilon  \oint_{\partial  \Sigma}  d \bar x \phi^{(0,\bar 1)}
|\psi
\rangle+\delta _\phi  | Q \psi\rangle}
where the contour integral over the boundary of $\Sigma$ comes from
the total
derivative term in \ope a - \ope d.
One can reinterpret this contour integral as coming from the
boundary
variation of the action that ensures
the BRST invariance of  the path integral on the
hemisphere~$\Sigma$.
The relation \change\ implies that  in the {\it  perturbed} theory
the BRST
operator depends on $\epsilon$ as follows
\eqn\defbrst{Q(\epsilon)= Q+ \epsilon \oint  \phi ^{(0,\bar1)},}
where $Q$ is the BRST operator of the unperturbed theory.
 There are  similar formulas for the other susy generators.
As one can see the variation of $G^- _0$ depends on $\bar   \epsilon$
and is
given as follows
\eqn\defb{G^- _0 (\bar \epsilon)=G^- _0+\bar \epsilon \oint d \bar z
{}~z ~
[\overline Q, \phi^{\dagger}]}
{}From \defbrst, it is clear that the contact term $\mu (\phi,\psi)$
is
not $Q$-closed. In fact, there should be $\bigl(Q+ \epsilon \oint
\phi
^{(1)}\bigr)\bigl(| \psi \rangle +\epsilon \,|\mu (\phi ,\psi
)\rangle \bigr)=
| Q\psi \rangle + \epsilon \,|\mu (\phi ,Q \psi )\rangle
\Longrightarrow $
\vskip 3mm

\eqnn\Nontriv
\hskip5pc \boxit{ \vbox{ \hsize 12pc
$$Q \mu (\phi ,\cdot) -  \mu (\phi , Q \cdot)=-\oint  \phi ^{(0, \bar
1)} $$}}
\hskip 7pc\raise1.2pc \hbox{\Nontriv}
\vskip 1mm

The variation of  the chiral states
\stateop\
induced by the contact
terms
gives rise to the (infinitesimal)
map of the Hilbert space $U_\phi : {\cal H} \rightarrow  {\cal H}$,
where $U_{\phi} |\psi \rangle =|\psi \rangle +\epsilon ^a |\mu
(\phi,\psi )\rangle$. This map  combines with \defbrst\  in a way
that ensures the
local background independence. Namely,
the representation of  the susy generators
changes according to
\eqna\obv{
$$\eqalignno{
U_\phi (Q+\epsilon \oint  \phi ^{(0,\bar 1)})~U_\phi ^{-1} | \psi
\rangle
&=\tilde Q
| \psi \rangle &\obv a \cr
  U_\phi (G^- _0+\bar \epsilon \oint d \bar z ~z ~ [\overline Q, \phi
^{\dagger}])~U_\phi ^{-1} | \psi \rangle &=\tilde G^- _0  | \psi
\rangle ~.
&\obv b \cr
}$$}
Similar formulas are valid for the right movers.
The operators $\tilde Q=Q + \delta Q$  and $\tilde G^- _0=G^- _0 +
\delta G^-
_0$ are the susy generators for the {\it unperturbed} theory
corresponding to
the point  $\tilde p\in {\cal M}$ close to the point $p$ where the
original
theory sits. Thus $ \delta Q$  and $\delta G^- _0$  are  the
classical
variations
due to explicit dependence of $N=2$ generators on the target space
metric and should not be confused with additional terms in \defbrst\
and \defb.

Operator $b^- _0= G^- _0 - {\overline G}^- _0$ plays the important
role
{\it in string theory}.
The Hilbert space ${\cal H}$ is defined as  follows
\eqn\hilbert{{\cal H}=\{ \psi \in {\cal H}| b^- _0 \psi=0 \}}
For $\bar \epsilon=0$ the map $U_{\phi}=1 + \epsilon \mu(\phi,
\cdot)$ defines
a flat connection on the
Hilbert space bundle
${\cal HM}$

\vskip 2mm
\eqnn\flatcon
\hskip6pc \Boxit{ \vbox{ \hsize 10pc
$$|  \psi \rangle  \rightarrow  |  \psi \rangle + \epsilon |
\mu(\phi, \psi
)\rangle $$}}
\hskip 8pc\raise1.2pc \hbox{\flatcon}
\vskip 2mm

\noindent
 Indeed, it follows from  \obv b\ that
$U_\phi b^- _0 U_\phi ^{-1} =\tilde b^- _0$  and therefore
$U_{\phi}$ maps  ${\cal H}_p$ on ${\cal H}_{\tilde p}$.
In the next section we will derive the semiclassical expression for
$\mu(\phi,
\cdot)$.

In fact  there are two equally good descriptions of the perturbed
theory. In
one description
a BRST operator $Q(\epsilon)$ varies according to \defbrst, while
$b^- _0$
remains
constant  (we already put $\bar \epsilon=0$, see \defb). There is
another
possibility to describe
the perturbed theory as an unperturbed one at the new background.
In this description the BRST operator  $Q$ and $b^- _0$ are mapped on
the
operators $\tilde Q$ and ${\tilde b}^- _0$ at the new background.
These two descriptions are equivalent to each other and
related by conjugation by operator $U_{\phi}$.

At this point one can use \contact\ to compute the effect of the
deformation
\action\ on the correlation functions. An important case is a
two-point
function of one chiral and one anti-chiral {\it primary} field which
gives a
matrix element of  the Zamolodchikov\foot{Depending on the
normalization,
it may be
either Zamolodchikov or $tt^*$ \ref\cv{S. Cecotti and C. Vafa,
Nucl.Phys. B367 (1991) 359-461} metric. The former is more natural
object for
the string is  theory, so we stick to it.} metric on $\cal M$:
$g_{b\bar
c}=\langle \phi _c^\dagger | \phi _b \rangle$.  We remind the reader
that the
primary fields are the ``harmonic" representatives of the BRST
cohomology. This
turns out to be quite important, because in the matrix element
$\langle \phi
_c^\dagger | \mu(\phi _a ,\phi _b)\rangle$ responsible for the chiral
deformation of  $g_{b\bar c}$, only the {\it harmonic} part of  the
contact
term contributes.  Indeed, taking the ``Hodge decomposition"
\ref\VLW{W.~Lerche, C.~Vafa, N.~Warner Nucl.Phys. B324 (1989) 427}~
\eqn\Hodstr{\mu (\phi _a,\phi _b)= \Gamma_{ab} ^c\,  \phi_c
+Q(...)+G_0(...)}
 one sees that  both $Q$- and $G_0$- exact terms decouple since
$\langle \phi
_c^\dagger |$ is harmonic: $\partial _a g_{b\bar c}=\Gamma _{ab}^c
g_{c\bar
c}$. In \Hodstr\ we decomposed the harmonic part
of the
contact term in the basis of  chiral primaries. By definition, the
coefficients
$\Gamma_{ab} ^c$ are the holomorphic Cristoffel symbols of  the
metric
connection $D_a$ for the Zamolodchikov metric \ref\kut{D. Kutasov,
Phys.
Lett. {\bf
220B} (1989) 153}.

A similar argument  can be applied to describe the deformation of
the
multipoint  correlation functions, possibly on the higher genus
worldsheet.
Indeed, in this case one integrates over the moduli space of
algebraic curves
with punctures.  The contact terms appear  on the compactification
divisors,
when two punctures try to collide.  They cannot possibly collide on
the
Deligne-Mamford compactified  moduli space \ref\DelM{P. Deligne and
D. Mumford,
Publ. Math. de l'IHES {\bf  36} (1969) 1}.
Instead,  a node develops on the curve by splitting off a
rational
curve with two ``colliding'' punctures on it.  In a language more
familiar to
physicists,  the colliding punctures sit on a sphere connected to the
rest of
the worldsheet by a long tube. The length (and twist) of the tube is
the right
parameter  to describe the closing of  the punctures.  In the limit
when the
punctures collide (``$z=x$ in the argument of the
$\delta$-function''),  the
tube is infinitely long.  Propagating along  the infinitely long
tube, the
state  $| \mu(\phi _a ,\phi _b)\rangle$  is automatically projected
onto the
ground states \cv.  In other words, again only the first term on the
right
hand side of \Hodstr\ happens to be relevant.

We see that  for the purposes of computing the correlation functions,
it
suffices to use the ``projected'' form of the deformation equations:
$\delta_a
 |  \phi_b \rangle = \Gamma _{ab}^c|\phi _c\rangle$ and $\delta_a
|
\phi_b^\dagger \rangle =0$. It describes the parallel transport of
the chiral
ground states with respect to  the Zamolodchikov connection $D_a$.
In fact one can make this projection more explicit.
Let us use the description of the perturbed theory
in which the BRST operator $Q(\epsilon)$ varies, while $b^- _0$
remains
constant.
The connection on the space of physical states should be compatible
with the
variations
of the BRST operator and $b^- _0$, namely
\item {$\bullet$} $Q$-closed states should be maped on
$Q(\epsilon)$-closed
states;
\item {$\bullet$} $Q$-exact states should be maped on
$Q(\epsilon)$-exact
states;
\item {$\bullet$} the variation of the states should be $b^- _0$
--trivial

\noindent
The first two conditions ensures that physical states are mapped on
the
physical states,
while the last condition ensures the uniqueness
of the connection \ref\los{A. Losev, private communication}.

Formally, for $|\psi \rangle$ being $Q$ -{\it closed} one can
immediately write
down
the connection that
satisfies the above conditions
\eqn\conlos{ | \psi \rangle  \rightarrow  |  \psi \rangle - \epsilon
{1 \over
Q} b^- _0  \phi  |  \psi \rangle ~.}
One can show that $b^- _0  \phi  |  \psi \rangle$ is always $Q$
exact and therefore $Q$ is invertible\foot
{This is an analog of $\partial \bar \partial$ Lemma for K\"ahler
manifolds.}.
The first two conditions follow from the identity
$\oint \phi^{(1)}   |  \psi \rangle  =  b^- _0  \phi  |  \psi
\rangle$ for
$\phi$ and $\psi$ being $Q$-closed.
The connection \conlos\
can be derived using the cancel propagator arguments \ref\dj{R.
Dijkgraaf,
talk at MIT; A. Losev, private communication}.

Describing the perturbed theory as an unperturbed one around the new
background
one obtains
the connection on the tangent bundle ${\cal TM}$

\vskip 5mm
\eqnn\project
\hskip5pc \boxit{ \vbox{ \hsize 14pc
$$|  \psi \rangle  \rightarrow  |  \psi \rangle +
\epsilon| \mu(\phi, \psi )\rangle- \epsilon {1 \over Q} b^- _0  \phi
|  \psi
\rangle ~.$$}}
\hskip 5pc\raise1.2pc \hbox{\project}
\vskip 1mm

\noindent
As we will see later both these connections \flatcon\ and \project\
appear in the description of AKS theory.

\subsec{Semiclassical calculations}

In the large volume limit the Hilbert space of the theory can be
identified
with the space of differential forms on the target space.
The left and right $U(1)$ charges can be  identified with
(holomorphic,
anti-holomorphic)
degree of the form. For this Hilbert space there is the following
dictionary
\eqn\dictionary{\eqalign{G^+ _0 \leftrightarrow \partial
{}~,&~~\overline G^+ _0
\leftrightarrow \overline \partial \cr
G^- _0\leftrightarrow \partial^{\dagger} ~,&~~\overline G^-  _0
\leftrightarrow
\overline \partial^{\dagger}  \cr} }
between the susy generators and the differential operators on $M$.
The total BRST operator $Q$ corresponds to $d$  while  $b^- _0= G^-
_0 -
{\overline G}^- _0$  corresponds to $d^{c \dagger}=
\partial^{\dagger}- \bar \partial ^{\dagger}$.
Let us also introduce the operator   $L$ of multiplication by
K\"ahler form
$\omega _{i\bar j}$ and the operator $\Lambda$ of contraction with
bivector
$\omega ^{i\bar j}$. The commutation relations between $L$,  $d^{c
\dagger}$
and $\Lambda$, $d$ are two of the Hodge identities
\eqn\hodgeid{[d, \Lambda]=d^{c \dagger} ~~{\rm and}~~~[d^{c \dagger},
L]=d}
(see  \ref\grhar{ P. Griffiths and J. Harris
{\it Principles of Algebraic Geometry} New York, Wiley, 1978} and
Appendix A).

Computing the OPE in
the $\sigma$-model formalism we arrive to the following formula:
\eqn\opecontact{{\cal O}^{(2)}_{\phi} (z) {\cal O}_{\psi}(x) \sim
...+
\delta ^2 (z-x) {\cal O}_{m(\phi,\psi)}(x)~,}
where $m(\phi ,\psi )$ is a bilinear symmetric  operation on
differential forms
defined by
\eqn\mab{
m(\phi,\psi)= \Lambda (\phi \wedge  \psi )-(\Lambda \phi )\wedge \psi
-\phi
\wedge
 ( \Lambda \psi )~.}
It has a degree ${\rm deg}m(\cdot, \cdot)=-2$.

For each K\"ahler form $\omega$, the linear operator $\Lambda$
descends on
cohomology $H^*(X)$. One should just  identify $H^*(X)$ with the
space of
harmonic forms.  It is easy to see that  the result  (denoted by the
same
symbol $\Lambda$) depends only on the {\it cohomological class}
$[\omega] \in
H^{1,1}$ of $\omega$.  Remarkably, we have a problem trying to use
the same
trick to descend the bilinear operation $m(\cdot, \cdot)$ on
cohomology.
Indeed, one can check that even for both $\phi$ and $\psi$ harmonic,
the
result  $m(\phi ,\psi )$ is not even $d$-closed.  The reason is that
the
product  $\phi \wedge \psi$ is not harmonic in general, so $d\Lambda
(\phi
\wedge \psi )=d^{c\dagger}(\phi \wedge \psi )\neq 0$. This is the
semiclassical
 manifestation of the relation \Nontriv. As it was discussed in the
previous
section,
we
introduce the operation $s(\cdot ,\cdot )$ on harmonic forms, defined
as
\eqn\sab{s(\phi ,\psi )=m(\phi ,\psi ) -{1 \over d}d^{c\dagger} \phi
\wedge
\psi, }
where  we have to use the Hodge decomposition of the product: $\phi
\wedge \psi
=h_{\phi \wedge \psi}+dc_{\phi \wedge \psi}$;
where harmonic   $h_{\phi \wedge \psi}$ satisfies
$d^{c \dagger} h_{\phi \wedge \psi}=0$.  The result of
$s(\cdot
,\cdot )$ is a $d$-closed form.  So defined,   $s(\cdot ,\cdot )$
descends to
cohomology (this will be proved in Section 4).  A reader can check
that  \sab\
is just a semiclassical version of  \Hodstr \foot{Similar
construction appears in topological B-models
as well as in topological LG theories. The  case of topological
B-model
will be discussed in Section 3.7.   For $\phi, \psi \in
H_{(0,\bar1)}(T_M)$ the
connection is given
$(\delta \psi)'=-\psi \perp \phi' -{1 \over \bar \partial}\partial
(\phi \wedge
\psi)'$, where
the contraction with holomorphic $3$ form is denoted by prime and
operation
$\perp$ is the contraction of  two holomorphic indices.
In the case of  topological LG theories the connection is given by
$\partial
(\phi \psi /W')_+$. Operation
$(\cdot /W' )_+$ is identified with $1/Q$, while $b_0 ^-$ -- with
$\partial$.}.

 Let us discuss the semi-classical (without instantons) $tt^*$ and
Zamolodchikov metrics on the complexified K\"ahler cone ${\rm K}_{\bf
C}\subset
H^{1,1}$.   It is convenient to  introduce a complexified K\"ahler
class
$\omega $ so that the positive definite real (true) K\"ahler  class
is $\Omega
=\omega +\bar \omega$.  We expect that in the absence of instantons
the metrics
depend only on $\Omega$. In the tangent space to K  there are
``chiral"
vectors  $\xi _a$ deforming  $\omega$ and their  ``antichiral"
counterparts
$\bar \xi _{\bar a}$ deforming  $\bar \omega$. Then the  (classical)
Zamolodchikov metric is defined by the  scalar product
\eqn\cltt{
\langle \bar \xi _{\bar a} \mid \xi _a \rangle={1\over {\rm
Vol}_\Omega} \int
(* \bar \xi _{\bar a} )\wedge \xi _a = {1\over {\rm Vol}_\Omega} \int
\Lambda
s(\bar \xi _{\bar a} , \xi _a)\Omega ^n=\Lambda s(\bar \xi _{\bar a}
, \xi
_a)~,
}
where both $\Lambda$ and $s(\cdot, \cdot)$ are computed with respect
to the
real class $\Omega$.
The $tt^*$ metric is given by the right hand side of \cltt\  without
${\rm
Vol}_\Omega ^{-1}$ prefactor.  It was noticed by Candelas
\ref\Ca{P. Candelas, Nucl. Phys. B329 (1990) 583}
(for general discussion see also \cv)
that  the metric \cltt\ is K\"ahlerian:
\eqn\Cand{
\langle \bar \xi _{\bar a} \mid \xi _a \rangle=\partial _a \bar
\partial _{\bar
a} \log{({\rm Vol}_\Omega  )^2}
}
The corresponding metric connection is
\eqn\Zacon{\eqalign{
D_a &=\partial _a-s(\xi _a, \cdot)\cr
 D_{\bar a} &=\bar \partial _{\bar a}-s(\bar \xi _{\bar a},\cdot)
}}
It satisfies $[D_a, D_b]=0$ and  $[D_{\bar a}, D_{\bar b}]=0$.  In
general, it
is not flat:  $[D_a, D_{\bar b}]\neq 0$.

The  simple example below may be helpful.  When  the cohomological
K\"ahler cone
K  is one-dimensional (generated by $x\in H^{1,1}(M)$),  the
complexified  cone
${\rm K}_{\bf C}$ is an upper half-plane of a complex parameter $z$.
The
K\"ahler form is given by $\Omega=2({\rm Im}\,z)\, x$ and the B field
by
$B=2({\rm Re}\,z)\, x$.  It is  easy  to compute the semiclassical
Zamolodchikov metric on ${\rm K}_{\bf C}$.  Using  \cltt\   one finds
that  it
is the Poincar\'e   metric
\eqn\Poinc{
g_{z\bar z}dzd\bar z=2n{dzd\bar z \over (z-\bar z)^2}=\partial \bar
\partial
\log{(z-\bar z)^{2n}},}
where $n={\rm dim}_{\bf C}M$.  The answer is essentially independent
of  any
detail of geometry of the  manifold $M$.  The   Zamolodchikov
connection is
given by
$\partial +2 (z-\bar z)^{-1}$ and $\bar \partial +2 (\bar z-
z)^{-1}$.  It has
constant negative curvature.

Now let us return to the general situation. Suppose that  $\phi$ is a
{\it
harmonic} $2$-form. The perturbation $\epsilon ^a\int  d^2 z~{\phi}_a
^{(2)}$
(~$\bar \epsilon ^a\int  d^2 z~[Q_-,[Q_+,{\phi} ^{\dagger}]]$~)
corresponds to
deformation of the K\"ahler form $\omega \rightarrow \tilde \omega
=\omega +
\epsilon ^a\phi _a$  (~$\bar \omega \rightarrow \tilde {\bar  \omega}
=\bar
\omega +  \bar \epsilon ^a\phi _a^\dagger$~).
Then it  follows from the formulas derived in  Appendix A that the
operators
$d$  and $d^{c \dagger}$ satisfy
\eqna\conss{
$$\eqalignno{
U_{\phi}^{-1}(d-\epsilon [d^{c\dagger},\phi ] )U_{\phi}  &=d &\conss
a\cr
U_{\phi}^{-1}(d^{c\dagger} -\bar \epsilon [d,\delta \Lambda
])U_{\phi}&=
(d^{c\dagger}-(\epsilon +\bar \epsilon )[d,\delta \Lambda ])~,&\conss
b \cr }$$}
where $U_{\phi}=1+\epsilon \, m(\phi,\cdot)+o(\epsilon ^2)$.
One can immediately recognize in these formulas a semiclassical limit
of \obv a - \obv b.
Indeed, the BRST operator $\tilde Q$  for the new background $\tilde
\Omega$ is
still $d$.
The operator  $\tilde G^- _0$ on the r.~h.~s.~ of \conss\ ~coincides
with
$d^{c \dagger}$ computed for $\tilde \Omega$ as a consequence of  the
Hodge
identities.

\subsec{$N=2$ CFT vs. $N=2$ Supersymmetric Quantum Mechanics.}
The two-dimensional sigma model has a  little brother --- the $N=2$
supersymmetric
quantum
mechanics (~1-d worldline sigma-model~). As a theory of topological
matter,
SQM is an approximation to the full-fledged 2-d sigma model. The
differnce
between them is that  SQM discards worldsheet instantons.

There are chiral and antichiral fields in the theory. It is
convenient to
identify the former with  differential forms on the target space and
the latter
with  polyvectors:
\eqn\qmstat{\eqalign{
\phi &=\phi_{i_1 ...i_n \bar j_1...\bar j_m} \chi^{i_1} ...
\chi^{i_n}
\overline \chi^{j_1} ... \overline \chi^{j_m}, \cr
\phi^\dagger &=\phi^{i_1 ...i_n \bar j_1...\bar j_m} \rho_{i_1} ...
\rho_{i_n}
\overline \rho_{j_1} ... \overline \rho_{j_m}. \cr
}}
The SQM susy generators are given by \dictionary.  The {\it chiral
primaries}
are harmonic forms, the  {\it antichiral primaries} can be obtained
by  raising
the indices of the latter.  Of course, up to now this  was just a
repetition of
the previous section.  The differences with  the string theory begin
when we
look at the space of deformations. There is no antisymmetric tensor
field
$B_{i\bar j}$ in SQM, so this theory is naturally defined on  the
{\it real}
K\"ahler cone K.  Our major assumption is that it possible to
identify the
deformations with real chiral fields. To explain this point,  let us
see how
the theory depends on the (real) K\"ahler form $\omega \in {\rm K}$.

As one moves along the K\"ahler cone, the space of harmonic forms (=
SQM ground
states~) changes.  It is explained in section 3.5 and  Appendix A
that the
natural parallel transport on harmonic forms is given by  a {\it
flat}
connection  $D_e^R=\partial _e -s(e,\cdot )$, where $e$ is any
tangent  vector
to the  real K\"ahler cone K, considered as a harmonic form.  It  is
also
explained in Appendix A, that  $D^R$ is not a metric connection with
respect to
the natural  (Hodge) metric  on K.  Let  us give a ``physical''
reason for
that.  We identified the tangent vectors to K with chiral fields.
In $N=2$ there is a
nondegenerate pairing  between chiral and
antichiral fields:
\eqn\veform{
\langle \phi ^\dagger |\psi \rangle ={1\over {\rm Vol}_\omega}\int
\phi ^{i\bar
j}\psi _{i\bar j}\, \omega ^n,}
(here $\omega$  is a real K\"ahler form).  Thus {\it a priory} it
defines a
pairing between
forms ans polyvectors but not a metric on K.
The connection $D^R$ preserves the pairing
\veform\ in
the following sense.  Let $\{e_a\}$ be the covariantly constant  with
respect
to $D^R$ coordinate frame  (~for each $\omega \in {\rm K}$,
$e_a(\omega)$ is a
harmonic form~).  Denote  by $\{\bar{e}_a\}$ the bivectors obtained
from
$\{e_a\}$ by raising the indices.  One can check\foot{Checking this
is a simple
yet good  exercise.} that so defined, $\{\bar{e}_a\}$ do not  depend
on
$\omega$:  $\partial _a\bar e_b=0$.  Then one finds
\eqn\conclpair{\partial _a \langle \bar e_b|e_c\rangle -\langle \bar
e_b|s(e_a,e_c)\rangle =\partial _a \langle \bar e_b|e_c\rangle
+\Gamma _{ac}^d
\langle \bar e_b|e_d\rangle =0.}

In fact there are two natural connections on the (real) K\"ahler
cone. One is
equal to
$D^R _e= \partial_e-s(e, \cdot)$,
while the second  -- is the {\it metric} connection and it is given
by
$D_e= \partial_e-{1 \over 2}s(e, \cdot)$.
The first one appears to be flat, while the second  one  does not.
A simple example from the previous section is rather helpful.
One dimensional (real) K\"ahler cone is equipped with a natural
metric
$(ds)^2={n \over 2}{(dy)^2 \over  y^2}$. The connection defined by
$s(\cdot,\cdot)$
is given by $D_e= \partial_e-2/y$, while the metric connection is
$D_e=
\partial_e-1/y$.

It is important to mention that $D_e= \partial_e-s(e, \cdot)$ is a
metric
connection on the {\it complexified} K\"ahler cone.
Comparing this with the previous section we conclude that  it is
indeed
possible to consistently identify the {\it real} ~$\omega$ in SQM
with the {\it
holomorphic} $\omega$ in $N=2$ TCFT and  the   ({\it real})
deformations of
SQM
with the {\it holomorphic} deformations of $N=2$ TCFT by means of
analytic
continuation.

\newsec{Theory of deformations of K\"ahler structures}

\subsec{Mirror for KS theory (AKS)}

The string field theory of topological B-model is related to
Kodaira-Spencer
theory \bcov, which describes
deformations of complex structure. It is natural to ask what is the
mirror of
this theory  \witcs.
The mirror of the string field theory clearly should be defined on
the loop
space ${\cal L} M$.  We will show below
that the semiclassical approximation (SQM) to this theory is related
to the
theory of
deformations of K\"ahler structures of $M$.
In the table
below we summarize the relations between deformations of K\"ahler
and
complex
structures that follow from
the comparison of the corresponding topological theories.

\bigskip
\hfil{}
\vbox{\tabskip=0pt \offinterlineskip
\def\tablerule{\noalign{\hrule}}
\halign to 250pt{\strut#& \vrule#\tabskip=1em plus 2em&
  \hfil#& \vrule#& \hfil#\hfil& \vrule#&
  \hfil#& \vrule#\tabskip=0pt\cr\tablerule
&&\omit\hidewidth  \hidewidth&&
  \omit\hidewidth B model (KS)\hidewidth&&
  \omit\hidewidth A model (AKS)\hidewidth&\cr\tablerule
&&\omit\hidewidth  BRST
\hidewidth&&$\overline{\partial}$&&$d~~~~~~$&\cr\tablerule
&&\omit\hidewidth  $b^-_0$\hidewidth&&$\partial$&&$d^{c
\dagger}~~~~$&\cr\tablerule
&&\omit\hidewidth   field\hidewidth&&$A \in \Omega^1 (T^*_M)$&&$K_0
\in
\Omega^{(1,1)}$&\cr\tablerule
&& constraint&&$\partial A'=0$&&
$d^{c \dagger}K_0=0$ &\cr\tablerule
\hfil\cr}}
\bigskip
\noindent
A Kodaira -Spencer equation is given by
\eqn\ks{\bar  \partial A'+ {1 \over 2}   \partial (A \wedge A)'=0~.}
Using the analogy with $B$ model one can write its mirror image as
\eqn\aks{dK_0 +{1 \over 2}  d^{c \dagger} (K_0 \wedge K_0)=0~,}
where $K_0$ satisfies a constraint $d^{c\dagger}K_0=0$.
Kodaira-Spencer equation \ks\ implies that  the deformed
BRST operator $\bar \partial + A \cdot \partial$ is nilpotent.
Similarly to B-model,  equation \aks\ is equivalent to the condition
that an
operator
$$D=d + [d^{c \dagger}, K_0]$$  squares to zero.

Now let us suppose that the manifold $M$ is 3-dimensional. We will
explain the
necessary modifications in the general case later when discussing the
BV
formalism.
  Then the equation \aks\ can be derived as an equation of motion for
the
action
\eqn\act{S[x,K]={1 \over 2 g^2}\int K {1 \over d^{c\dagger}} dK   +
{1
\over
6 g^2}\int (x+K) \wedge (x+K) \wedge (x+K)~.}
In \act, we separate the contributions of massless and massive modes.
We call $x
\in {\rm Ker}d \cap {\rm Ker}d^{c\dagger}$ massless and
$K=d^{c\dagger}Z$
massive. There is an  ambiguity in this definition. On  the one
hand, one can
show that massless modes can be parameterized as
$x=h_x+d d^{c \dagger }\beta $, where $h_x$ is harmonic \foot{this
decomposition requires a choice of complex structure} (see Appendix
A).
On the other hand, the massive mode $K$ in \act\ is not  defined
canonically
either, since a shift  of $K$ by $dd^{c\dagger}\beta $ does not
affect the
kinetic term. Finally, a formula  $S[x,K] =S[h_x, K+
dd^{c\dagger}\beta ]$
shows that one may always fix $x$ to be a harmonic 2-form,
once a complex structure is chosen.  Below we adhere to this
interpretation of
the massless mode.

The massive mode $K$ is the dynamical variable in the theory, while
the
massless mode
plays the r\^ole of  background. The action  \act\  possesses gauge
invariance
discussed below.  After imposing a {\it gauge fixing} condition
$d^\dagger K=0$
 a  propagator for the massive modes can be written as
\eqn\prop{\Pi=d^{\dagger} {1 \over \Delta} d^{c\dagger}~.}

Having  defined the propagator $\Pi$ we can rewrite the formula \sab\
as
follows
\eqn\decomp{s(\phi ,\psi )=P_{{\rm Ker} (d)} m(\phi ,\psi )=m(\phi
,\psi ) -
\Pi(\phi \wedge \psi )~.}
In the target space field theory such as AKS the massless modes do
not
propagate. On the contrary, in the string theory the massless modes
are
physical and do propagate.
The propagator of the massless modes $x$ is related to the connection
\decomp\
and is given as $D(\phi, \psi)=s(\phi, \psi)+ (\Lambda \phi) \psi +
\phi
(\Lambda \psi)$.
It is clear that $D(\phi, \psi)$ depends only on the product $\phi
\wedge
\psi$, but not on $\psi$
and $\phi$ separately.

After imposing the gauge fixing condition $d ^{\dagger}K_0=0$ one can
solve
\aks\ in perturbation series.
To write the solution one needs to fix the harmonic part  $x$ of
$K_0$ (in the
sense of the Hodge decomposition  $K_0=x+dN$).
The perturbation series for  \aks\ formally coincides with the
perturbation
series of  $\phi^3$ theory
coupled to the background:
\eqn\Sol{K_0[x]=x-{1 \over 2}\Pi ({x}^{2})-{1 \over 2}\Pi (x\Pi
(x^2))+\cdots~.}
We see that  (the gauge-fixed) solution of  the equation of motion is
completely
determined by the massless mode $x$. In other words, the (physical)
covariant
phase space consists of  harmonic  $(1,1)$-forms.

The solution $K_0[x]$ defines a deformed K\"ahler structure  in the
same way as
the solution of Kodaira-Spencer equation defines a new complex
structure.
In fact (see section 4),  for any $x$ there exist a new K\"ahler
structure
$\tilde \omega (x)$ and an operator $U$,
such that
\eqn\back{\eqalign{D=d+[d^{c \dagger}, K_0[x]]=UdU^{-1}  \cr
d^{c \dagger}=U {\tilde d}^{c \dagger} U^{-1}~.\cr}}
This is an analog, for deformations of K\"ahler structures, of
Tian-Todorov
\ref\tian{ G. Tian, in {\it Essays on Mirror manifolds, ed. by
S.~T.~Yau},
International Press, 1992}
\ref\tianI{G. Tian, in {\it Mathematical aspects of String theory,
ed. by S.T.
Yau}, World Scientific, Singapore, 1987},  \ref\Tod{ A.N. Todorov,
Comm. Math.
Phys. 126 (1989) 325}  construction.

The relations \back\ express the global background independence of
AKS the same
way as \conss~ express the local background independence of  string
theory.
There is an obvious difference between these two theories.  The
former  (AKS)
is defined on the tangent bundle ${\cal T}{\rm K}$ to the {\it real}
K\"ahler
cone K ($\omega$ gives a point on the base and $x$ a vector in the
tangent
space).  The latter  (perturbed string theory, see section 2 ) is
defined  on
the tangent bundle ${\cal T}{\rm K}_{\bf C}$ to the {\it
complexified} K\"ahler
cone ${\rm K}_{\bf C}$.
In fact, it is natural  to interpret  AKS as a target space field
theory
for
$N=2$ susy quantum mechanics\foot{We are grateful to C.~Vafa for this
suggestion.}.  The latter is a semiclassical approximation to $N=2$
CFT, in a
sense specified in 3.1.

This view  on AKS  as a large volume string field theory is supported
by  a
number of properties it  enjoys.  This theory is
\item {$\bullet$}   gauge invariant (for $3$-fold the gauge group is
the group
of volume preserving diffeomorphisms);
\item {$\bullet$}    independent of the complex structure;
\item {$\bullet$}  depends only on K\"ahler {\it class} of the
metric;
\item {$\bullet$}   background independent.

\subsec{Independence of complex structure}

We expect that  AKS, as  a target space field theory for the A type
$\sigma$-model,  is independent of the complex  structure, for the
fixed
K\"ahler form $\omega$. The metrics $g_{ab}$, the complex structure
$J^a_{\ b}$
and $\omega$ are related by
\eqn\trivrel{
g_{ab}=J^c_{\ a}\omega _{cb}}
In particular it means that as we change the complex structure with
$\omega$
fixed, we change the metrics and consequently the operators
$d^{\dagger}$ and
$\Delta$.
Now the  point is that  the particular combination of    $g_{ab}$ and
$J^a_{\
b}$  in $d^{c\dagger}$ results in dependence of   $d^{c\dagger}$ on
just
$\omega$, as can be seen from the formula
\eqn\trivrelI{
d^{c\dagger}=[d, \Lambda]}
where $\Lambda$ is a bivector such that $\Lambda ^{ab}\omega
_{bc}=\delta ^a
_c$. Therefore both the kinetic term of  \act\ and the constraint
$K=d^{c\dagger} Z$ are  independent of  $J^a_{\ b}$.

The only subtlety is that the Hodge decomposition  which we use to
choose the
harmonic representative for $x$, depends on the complex structure.
But  as the
complex structure changes, $x$ changes by a $d$-exact form, which
does not
affect the kinetic term and can be reabsorbed into the massive $K$.

\subsec{ Gauge invariance}
 Action   \act\
is  invariant under  the gauge transformation
\eqn\gauge{\delta_{\alpha}K= d\alpha-d^{c\dagger}((x+K) \wedge \alpha
)~,}
where $\alpha$ is an infinitesimal form such that $d^{c\dagger}\alpha
=0$. Note
that only the massive mode $K$ gets transformed\foot{Using the
results of
Appendix A one can easily show that  the variation \gauge\ is
massive: $\delta
K=d^{c\dagger}w$} leaving the background field $x$ unchanged.

Indeed,  keeping only the linear terms the variation can be written
as
\eqna\brstinv{
$$\eqalignno{
g^2 \delta S&=\int -K {1 \over d^{c\dagger}}d d^{c\dagger}((x+K)
\wedge
\alpha) +
{1\over 2}\bigl[d\alpha-d^{c\dagger}((x+K) \wedge \alpha)\bigr]\wedge
(x+K)^2   & \cr
&=\int -{1\over 2}d[(K+x)^2] \wedge \alpha + {1\over
2}\bigl[d\alpha]\wedge (x+K)^2 + {1\over 6}\alpha
d^{c\dagger}[(x+K)^3]& \cr
&=\int -{1\over 6}d^{c\dagger}[\alpha ] (x+K)^3 =0,\cr}$$
}
where we used  that  $d^{c\dagger}$ is antihermitean and  a formula
$Ad^{c\dagger}( A^2)={1\over 3}(A^3)$
 which holds if $d^{c\dagger}A=0$.
The commutator of two gauge transformations is given as follows
\eqn\comm{[\delta_{\alpha},\delta_{\beta} ]=\delta_{\gamma}~,~~~{\rm
where}~~~
\gamma=d^{\dagger c}  (\alpha \wedge \beta)}
Define a correspondence $\alpha \leftrightarrow \check{\alpha} $
between the
the gauge parameters and the volume preserving vector fields by the
formula
$\check{\alpha}^I= \Lambda^{I J}   \alpha_J$.
Then \comm\  is equivalent to the commutation relation
$[\check{\alpha},\check{\beta}]=\check{\gamma} $ of the corresponding
vector
fields.
In the case of $3$-fold  this implies that the gauge group
is isomorphic to  the group $SDiff\, M$ of  volume preserving
diffeomorphisms.

Let us define field strength as
$F=dK+ {1 \over 2} d^{c \dagger} ((x+K)^2)$.
The condition of flat connection ($F=0$)  implies that $D=d+
[d^{c \dagger} ,(x+K)]$ is nilpotent.
The field strength $F$ is not invariant under gauge transformation
as it supposed to be in non-abelian
gauge theories
and transforms as follows
$ \delta F =   -d^{c \dagger}  (F \wedge \alpha)$.

\subsec{BV quantization of AKS.}

Our aim is to establish AKS as a target space field theory for $N=2$
SQM. But
the theory described so far  is a one describing only 2-forms, while
the
states \qmstat\ of   SQM are differential forms of  {\it all}
possible degrees.
Thus the ``superparticle field"  should rather  be a linear
combination ${\cal
K}=\sum_{q=0}^n K_q$,  each component  describing the sector  with a
particular
ghost number\foot{In this section we use both terms ``the ghost
number" and
``degree of form'' with the same meaning. }.  Also, as mentioned
above, the
action \act\  works only for 3-dimensional $M$.  In fact, these two
problems
turn out to be each other's cure. Adding extra fields corresponding
to all
degrees of freedom of  SQM also makes the theory well defined for any
dimensional $M$.

 But we don't  even have to appeal  to any {\it a fortiorti}
connection to SQM or to the case ${\rm dim}\, M\neq 3$.  A
consistent
treatment of the theory with action \act\ within the
Batalin--Vilkovisky
(BV) formalism \ref\bat{ I. A. Batalin and G. A. Vilkovisky,
Phys. Rev   D28  (1983) 2567} (for review see also \ref\hen{ M.
Henneaux, {\it
Lectures on the antifield-BRST formalism for gauge theories}, Proc.
of XXII
GIFT Meeting} and in string theory  \ref\zw{B. Zwiebach, {\it Closed
String
Field Theory: Quantum Action and the B-V Master equation},
IASSNS-HEP-92/41, MIT-CTP-2102, hep-th/9206084})) requires one to
relax the
condition that $K$ is a 2-form and
includes all possible fields with arbitrary ghost numbers.  The
components $K_q
\in  \Omega ^{q} (M)$ with ghost numbers $q(K) \leq n-1$ are called
fields,
while the  components $K_q^*\in \Omega ^{q} (M)$ with ghost numbers
$q(K) >
n-1$ are called
antifields.  Both fields and antifields satisfy the constraint
$d^{c\dagger}K_q=0$ and can be decomposed  into the sum of  massive
and
massless
modes. The massless modes $x_q$ are the harmonic forms on $M$. They
are not
dynamical and just create the background.  The massive modes (from
now on
denoted by $K_q$ ) are dynamical. They  satisfy
$K_q=d^{c\dagger}Z_{q+1}$. Note
that the last condition implies there is no  dynamical  (anti)field
$K_{2n}$ of
the highest rank.     Thus there is the same number  (n-1) of fields
and
antifields.

The space of  of fields--antifields is equipped with an odd
antibracket\foot{In fact, this antibracket is induced by a natural
(even)
Poisson bracket on differential forms
$\{ K_p (z), Z_{q} (w) \}_ {_{\rm P.B.}} =
 \delta_{p+q,2n}\, \omega^{n}\delta(z,w)~. $ The latter will be used
to define
a measure in the path integral formulation of AKS (see Section 5).}
$\{ ~,~ \}$  given
by
$$\{ K_p (z), K^* _{q}  (w) \} =
 \delta_{p+q,2n-1}\, \omega^{n} d^{c\dagger}\delta(z,w)~,$$
where $\delta (z,w)$ is the delta function on
the K\"ahler manifold, such that for any {\it function}  $\varphi
(x)$
$$\int_M \varphi (y)\delta (x,y) \omega ^n(x)=\varphi (x)~.$$
It pairs $\Omega ^p(M)$ with  $\Omega ^{2n-p-1}(M)$.
This structure is promoted to a
canonical antibracket on the space of functionals:
\eqn\bvbrac{\lbrace{ F,L \rbrace}=
\int  \sum_{p+q=2n-1}   d^{c\dagger} \left(  {\delta F \over \delta
K_p}
\right)
 {\delta L \over \delta K_q^{}} -
{\delta F \over \delta K_q^{}}
 d^{c\dagger} \left(
 {\delta L \over \delta K_p}\right)  ~.}

In general BV theory, the BRST symmetry is a canonical transformation
in the
antibracket:
\eqn\bvbrst{\delta_{BRST} {\cal K}=\lbrace{ {\cal K},S \rbrace}~,}
where the original action (\act\ in our case) is replaced by a full
action $ S$
which
depends on both fields and antifields.  The full action satisfies two
conditions.
It   reduces to the original action when all antifields are
set to zero. It also
satisfies a Batalin-Vilkovisky master equation
\eqn\bvq{
\lbrace{ S,S \rbrace}=\hbar \Delta S~,}
where $\Delta$ is the natural Laplacian on the space of
fields--antifields to be defined below.  The r.h.s. of \bvq\ is a
contribution
coming from the path
integral measure.
At the classical level $(\hbar=0)$, the Batalin-Vilkovisky equation
is
nothing else but the condition that  the  full action is gauge
invariant.
The gauged fixed action is determined by
an odd functional $\Psi(K)$ and is given
by $S_{\Psi}(K)=S(K,K^*=\delta \Psi / \delta K)$.

It is quite remarkable that the full AKS action is given by the same
expression \act\ as the original AKS action, but without any
restrictions on
the
ghost numbers. One should simply substitute in \act\  the linear
combination
$\cal K$ for $K$:
\eqn\fact{ S[x,{\cal K}]={1 \over 2 g^2}\int
{\cal K} {1\over d^{c\dagger}}d{\cal K}   +  {1 \over 6 g^2}\int    (
x+{\cal
K})
\wedge (x+{\cal K}) \wedge (x+{\cal K})~,}
\noindent
or written in components
$${1 \over 2 g^2}\int \sum_{p+q=2n-2} K_p {1 \over d^{c\dagger}}
dK_{q}
+  {1
\over 6 g^2}\int \sum_{p+q+r=2n}(x_p+K_p) \wedge (x_q+K_q) \wedge
(x_r+K_r)~.$$

 To see why this is true we first notice that  if  M is
3-dimensional, every
term in \fact\  either consists of  2-forms or contains at least one
antifield
(form of rank $>2$).
When all antifields are set to zero, the only contribution to the
action comes
from the original field $K_2 \in \Omega ^{2} (M)$.
Now let us consider the BRST symmetry \bvbrst\ generated by \fact\
together
with \bvbrac.  One easily finds
\eqn\trank{\delta _{BRST}{\cal K}=d{\cal K} +{1\over
2}d^{c\dagger}(({\cal
K}+x)\wedge({\cal K}+x)).}
For 3-dimensional $M$  this formula with antifields set to zero
brings us back
to  \gauge, where the parameter $\alpha$ is to be identified with
$K_1+x_1$.

Now we can check the gauge invariance of  the full action \fact. The
computation itself  mostly repeats the one \brstinv\  ~done in the
previous
section.  It gives $\delta _{BRST}S=\{S,S\}=0$.

Note that the right hand side of  the BRST transformation \trank\
coincides
with  equations of motion for the action \fact.  BRST triviality of
the
dynamical equations  may not be surprising  after all, if  we notice
that  the
action \fact\ can also be written in a form
\eqn\factI{S[x,{\cal K}]={1 \over 2 g^2}\int {\cal K}\wedge \omega
\wedge
{\cal K}
 +  {1 \over 6 g^2}\int    ( x+{\cal K}) \wedge (x+{\cal K}) \wedge
(x+{\cal K})~,}
particularly useful in applications (we used the Hodge identity
\hodgeid\ and
the constraint ${\cal K}=d^{c\dagger}{\cal Z}$).  After gauge fixing
the
solutions of equations of motion can be expressed in terms of the
massless
modes $x_q$ by series similar  to \Sol.  Therefore the  BV {\it
covariant phase
space}, alias the space of solutions  coincides with the space of
harmonic forms
modulo the gauge group.  In particular, it is finite-dimensional.

The BV Laplacian is defined by:
$$\Delta =\int \sum_{p+q=2n-1}{\delta \over \delta K_q} d^{c\dagger}
{\delta
\over \delta K_p}~.$$

To verify that this definition is indeed covariant one has to take
into account
that
$\delta K_p  (x)/ \delta K_r  (y)=\delta_{p,r} \delta (x,y) \omega ^n
(x) $.
Now we can check  the BV master equation \bvq.
The gauge invariance of the full action implies that l.h.s of
\bvq\ is equal to zero. The r.h.s can be  computed easily.  It
 equals
$$ \Delta S  = {1 \over g^2} \int d^{c\dagger}(K_1 +x_1)\wedge
\omega^n(x) =0$$
 due to
constraint.

The above discussion implies that quantum corrections
are not needed for maintaining the gauge invariance of the
AKS theory.  There is no 4- or higher  interaction vertices in the
full action
\fact.  The same situation was encountered in
\ref\Konts{M.~Kontsevich, Talk at Princeton University}
and \witcs\  describing Chern-Simons
theory.
 These facts have similar geometric reasons.  The same reasons
guarantee
existence of series \Sol\  for AKS (and similar series for  CS
theory)
describing the solution of the equation of motion in terms of
massless
component $x$.  The higher vertices are  related  \witcs\ to the
higher {\it
Massey products} in cohomology.  The nontrivial Massey products are
obstructions to writing formulas like \Sol\ since $d$ cannot be
inverted.  In
the cohomological theory which appears in CS  (``cohomology with
coefficients
in $End(E)$") the Massey products are absent.  A very important fact
about
topology of K\"ahler manifolds is that  there are no Massey higher
products
either  --- it is a consequence of  $dd^c$--lemma (see
\ref\MGD{P.~Deligne,
P.~Griffiths, J.~Morgan,  D.~Sullivan {\it Real Homotopy Theory of
K\"ahler
Manifolds} Invent.~Math.~{\bf 29}, 245-274 (1975)} for  a proof  and
a nice
exposition on important consequences of this fact).
\subsec{ K\"ahler  ``topological" invariance.}

The action \act~is invariant under the variation of the K\"ahler form
$\omega
\rightarrow \omega+d \alpha$.
 To prove this, it is convenient to use another form of the action
\act
\eqn\actI{S(K)={1 \over 2 g^2}\int K \omega K   +  {1 \over 6
g^2}\int
(K+x)^3,~}
 equivalent to \act\  on the constraint $K=d^{c\dagger}Z$.
In \actI, the variation of  the kinetic term is due to variation both
of the
K\"ahler form  and the field $K$. The field $K$ changes because the
constraint
$K=d^{c\dagger}Z$ explicitly depends on the metrics:
\eqna\varl{
$$\eqalignno{
\delta d^{c\dagger}&=[d^{c\dagger},[\Lambda ,d \alpha ]]  &\varl a
\cr
(K+\delta K)&=(d^{c\dagger}+\delta d^{c\dagger})(Z+\delta Z) &\varl b
\cr
\delta K&=-[\Lambda ,d \alpha ]K + d^{c\dagger}\chi &\varl c \cr
 }$$}
where we choose $\chi=\alpha K$. In this case one can rewrite the
variation
as $ \delta K={\cal L}_\xi K$,
where  ${\cal L}_\xi K$ is the Lie derivative along the vector field
$\xi$
dual to $1$-form $\alpha$.
This is a clever choice since the variation of the {\it harmonic}
part $x$ is
also given by Lie derivative
$\delta x ={\cal L}_\xi x$.
Therefore, if we use $\xi=\alpha K$, we have a natural relation
\eqn\nat{
\delta (K+x)={\cal L}_\xi(K+x)
}

The variation of the kinetic term is
\eqn\varacI{\eqalign{
\delta _{\rm Kin} S=&{1 \over 2 g^2}\int K d \alpha K +2({\cal L}_\xi
K)
\omega K \cr
=&{1 \over 2 g^2}\int  K d \alpha K + {\cal L}_\xi (K\omega
K)-K({\cal
L}_\xi
\omega )K=0 \cr
}}
since
${\cal L}_\xi  \omega  =d(i(\xi )\omega)=d\alpha $
and
\eqn\zerid{
\int {\cal L}_\xi (K\omega K)=\int (di(\xi )+i(\xi )d)(K\wedge \omega
\wedge
K)=0
}
as $K\wedge \omega \wedge K$ is a top form.

The variation of the potential term is
\eqn\varacII{
\delta _{\rm Pot} S={1 \over 2 g^2}\int  ({\cal L}_\xi
(K+x))(K+x)^2={1
\over
6 g^2}\int  {\cal L}_\xi (K+x)^3=0
}
where we used \nat\ and the same argument as in \zerid.

\subsec{ Dependence on the K\"ahler class.}

To define AKS theory one needs to fix some data ---  K\"ahler
structure $\omega$
and massless background $x$.
It turns out that this data is redundant. AKS action possess
additional
symmetry which acts on the background data
$\omega \rightarrow \tilde \omega$ and $x \rightarrow \tilde x
(\tilde \omega)$
as well as on
$K \rightarrow \tilde K(\tilde \omega)$ such that  the action \act\
is almost
invariant, namely

\vskip3mm
\eqnn\efac
\hskip0pc \boxit{\vbox{\hsize 22pc
$${1 \over   {\rm Vol}^2 _{ \omega}}\bigl( S[ x,K;\,  \omega ]- S_0[
x ;\,
\omega] \bigr)= {1 \over {\rm Vol}^2 _{\tilde \omega}}\bigl( S[\tilde
x,\tilde
K;\, \tilde \omega ] - S_0[\tilde x ;\, \tilde \omega] \big)$$}}
\hskip0pc\raise1pc\vbox{\efac}

\noindent
where  $S_0 [x,\omega]$ is the classical action evaluated on the
solution $K_0
[x]$ and functions $x (\tilde \omega)$,
$K(\tilde \omega)$ satisfy differential equations discussed below.
The
combination that appeared in \efac\
may be viewed as a background independent action. The second term
does not
depend on the dynamical variable $K$  and therefore
does not affect the equations of motion. The appearance of the volume
factor in
front  of the action is quite remarkable and can be viewed as  volume
dependence of  the string coupling constant.

To prove \efac\ let us consider an infinitesimal variation
$\omega \rightarrow  \omega +\delta \omega$ by  {\it harmonic} form
$\delta
\omega$ accompanied by the following transformation of fields
\eqna\newfield{
$$\eqalignno{
&x \rightarrow \tilde x= x + \delta \omega-s(\delta \omega,x)
&\newfield a\cr
&K \rightarrow \tilde K  = K -m(\delta \omega, K+x)+s(\delta
\omega,x) ~.
&\newfield b\cr
}$$}
One can check that  the deformed fields satisfy
the constraint ${\tilde d}^{c \dagger} \tilde x=0$ and $d\tilde x=0$
as well as
 $\tilde K={\tilde d}^{c \dagger}  Z$. Therefore the transformation
\newfield~~is consistent with the decomposition on massless and
massive modes.
Moreover,  \newfield~~preserves the gauge fixing: $d^\dagger x=\tilde
d^\dagger
\tilde x=0$.

As the K\"ahler form changes by an infinitesimal {\it harmonic} form
$\delta
\omega$, the action \act\ changes so that
\eqn\ktsp{\eqalign{
g^2 S[&\tilde K,\tilde x;\, \omega+ \delta \omega]-g^2 S[ K, x;\,
\omega  ]=
\cr
&=  \int {1\over 2} K\delta \omega K -m(\delta \omega,K)\omega K
-\bigl[-s(x,\delta \omega ) +m(\delta \omega,x)\bigr]\omega K +\cr
&+ \int   {1\over 2}\delta \omega (K+x)^2- {1\over 6}m\bigl(\delta
\omega,(K+x)^3\bigr) =\cr
&= - \int m \bigl(\delta \omega, {1\over 2}  K \omega K +{1\over
6}(K+
x)^3\bigr)+
{1 \over 2}  \int \delta \omega x^2  \cr
&+ \int [\delta \omega K x-\bigl(-s(x,\delta \omega) +m(\delta
\omega,x)\bigr)\omega K]
\cr
&=2 (\Lambda \delta \omega) g^2 S[ K, x;\, \omega  ] + {  1 \over 2}
\int
\delta
\omega x^2 \cr
}}

This derivation deserves a few comments. First, we used the fact that
$m(\cdot,
\cdot)$ {\it differentiates} multiplication of forms (see Appendix
A).
Second, $(\Lambda \delta \omega)$ is a {\it number} since
$d^{c\dagger} \delta
\omega =0$ and therefore
$$\int m(\delta \omega ,{\cal L})=-2(\Lambda \delta \omega) \int
{\cal L}~.$$
Third, to obtain the last identity we noticed that since
$K=d^{c\dagger}Z$  and $m(\delta \omega,x) -s(x,\delta \omega )
=\Pi(\delta
\omega x)$, where $\Pi$ is the massive propagator
(see \decomp\ ) one can write
$$
\int K\omega\bigl(-s(x,\delta \omega ) +m(\delta \omega,x)\bigr)=
\int Z[d^{c\dagger},\omega ]  \Pi(\delta \omega x)=\int Zd \Pi
(\delta \omega
x)=\int K\delta \omega x.$$

One can see that the variation of the action consists of two terms;
the first
one is just  the rescaled action, while the
second is $K$ independent.  Therefore the {\it difference} $S[K,x;\
\omega]-S[Q,x;\ \omega]$  scales  by a factor $(1+2 (\Lambda \delta
\omega))$
under the  transformation \newfield\ . It is important that  the
solution
$K_0[x]$ of the equation of motion \aks\
 is mapped on the solution of  the equation of motion for the
perturbed
K\"ahler structure $\tilde \omega$ which can be written
as $K_0[\tilde x ]$. Taking $Q=K_0[x]$
we conclude that $S[K,x;\ \omega]-S_0 [x;\ \omega]$ scales under the
variation
\newfield\ .
The scale factor $(1+2 (\Lambda \delta \omega))$  is related to the
variation
of the of volume as follows
$$\delta {\rm Vol}_{\omega}=3 \int \omega^2 \delta \omega =
                 - {1 \over 2} \int m(\delta \omega, \omega
^3)=(\Lambda
\delta
\omega) {\rm Vol}_{\omega}~.$$
This implies the infinitesimal form of  the relation \efac.

 We will call a combination that appears in \efac\ the background
independent
action\foot{It is worth mentioning that  it is $S[K,x;\ \omega]/{\rm
Vol}^2
_{\omega}$,  not
${\cal A}[K,x;\ \omega]$ which
is related to the generating function of string amplitudes.}
\eqn\backind{\eqalign{{\cal A} &[K', x ;~ \omega]= {1 \over   {\rm
Vol}^2 _{
\omega}}\bigl( S[ x,K;\,  \omega ]- S_0[ x ;\,  \omega] \bigr)=\cr
 &= {1 \over g^2  {\rm Vol}^2 _{ \omega}}  \int \Bigl({1 \over 2} K'
{1
\over
d^{c\dagger}} D K'   +  {1 \over 6} K' \wedge K' \wedge K' \Bigr) ~,
\cr }}
where   we introduce a new dynamical variable $K'=x+K-K_0$ shifting
$K$
by the
classical solution  $K_0$. The differential operator $D$ is given by
\eqn\difop{D=d+[d^{c \dagger}, K_0]~.}
A transformation \newfield\ ~is a symmetry of  background independent
action
${\cal A} [K', x ; ~\omega]$.  Written as
\eqn\parallel{D_a x + \partial_a \omega=0~,~~{\rm
where}~~~D_a=\partial_a +
s(\partial_a \omega, \cdot)}
 \newfield a\ defines the parallel transport on the space of massless
modes.
The properties of this system will be discussed in the next section.
Similarly
one can define the parallel transport $K'(\omega)$ of $K'$.
Consider the solution  $ x(\tilde \omega)$ of   \parallel, satisfying
the
initial condition $x(\omega)=x$.
The action \backind\  evaluated on $ x(\tilde \omega)$ and $K'(\tilde
\omega )$
is independent of $\tilde \omega$.

Under an infinitesimal change of K\"ahler structure $\omega
\rightarrow \omega
+ \delta \omega$ the solution \Sol\ of  the equation of motion
transforms as
follows: $K_0 \rightarrow \tilde K_0=K_0+\delta \omega-  m(\delta
\omega,K_0)$.
 It is easy to convince oneself that $\tilde K_0$ satisfies the
equation of
motion with respect to the {\it new} background:
$$d \tilde K_0 +{1 \over 2}  {\tilde d}^{c \dagger} ( \tilde K_0
\wedge  \tilde
K_0)=0~ ~{\rm and}~~~{\tilde d}^{c \dagger}  \tilde K_0=0~.$$
At the same time the operators   $D$  and $d^{c \dagger}$ transform
by
conjugation  by  the operator $U_{\delta \omega}=1+m(\delta \omega,
\cdot)+o(\delta \omega ^2)$ as follows:
\eqn\conn{\eqalign{
D \rightarrow U_{\delta \omega}^{-1}DU_{\delta \omega}  &=\tilde D
\cr
d^{c\dagger}   \rightarrow  U_{\delta \omega}^{-1}d^{c\dagger}
U_{\delta
\omega}&=d^{c\dagger}+[d,\delta \Lambda ]={\tilde d}^{c\dagger} .
}}
As a result  the perturbed $\tilde D$ is given by $\tilde
D=d+[{\tilde
d}^{c\dagger}, \tilde K_0 ]$. We see that the symmetry \newfield\ ~
preserves
the relation between $D$ and $K_0$.

 In Section 4 we will show that  for each $x$ one can find $\omega_0$
such that
$\tilde x (\omega_0)=0  \Longrightarrow K_0=0$ and therefore $D=d$ in
that
background.  In this case the ``statement of background independence"
\efac\
can be written in a form familiar from \bcov.

The volume dependence in \backind\ deserves a  separate discussion.
If  one
introduces the volume-dependent  ``running string coupling constant''
$g_{\omega}$
which governs the magnitude of the cubic interaction, from \backind\
it follows
that

\vskip3mm
\eqnn\strcoco
\hskip9pc {\Boxit{\vbox{\hsize 6pc $$g_{\omega} ^2=g_0^2\,V^2$$}}}
\hskip9pc
\raise1pc
\hbox{\strcoco}

The reason for growth of  $g_{\omega}$ with volume $V$ is quite
clear.
For small
enough $V$, SQM is strongly interacting.  On the other
hand, the
$V\longrightarrow  \infty$ limit  for the {\it fixed} $g$ corresponds
to free
theory.  Background independence means that  the theory is the same
for all
values of $V$, therefore we should keep increasing $g$ as
$V\longrightarrow
\infty$ in order to preserve the nontriviality of cubic interaction.

What is really interesting in \strcoco\ is the parabolic rather then
linear
growth of  $g(V)^2$.  It suggests that the  field  $K'$ in  \backind\
scales
as $K'\rightarrow \lambda ^2 K'$ as the K\"ahler form $\omega$  goes
to
$\lambda \omega$. To understand this better, let us notice that  the
scaling
of $K'$ in \backind\ should coincide with that of the operator
$D\over
d^{c\dagger}$ in the kinetic term.  It is obvious that
$d^{c\dagger}\rightarrow
\lambda ^{-1}d^{c\dagger}$; the point is that  the BRST operator  $D$
also
changes: $D\rightarrow \lambda D$. The overall factor $\lambda ^2$ is
in
agreement with what we expect from \strcoco.

\subsec{KS theory and dependence on the complex structure}

This section probably would be more appropriate for \bcov.
Here we would like to discuss for KS theory a relation similar to
\efac, which
is a local
form of background independence.
We remind the reader that the basic field $Y$ in KS theory is a
$(0,1)$ form
with coefficients in
vector fields. The dynamical field in KS theory is not
$Y$ but its massive component $A$, while the massless component $x$
(cohomology
element)
plays the role of the
background.  For a fixed complex structure $J$ and a cohomology
element
$x \in H^{(0,1)}(T_M)$ the KS action reads as follows
\eqn\ksaction{S_{KS}[A,x;J]={1 \over 2}  \int A' { 1\over \partial}
\bar
\partial A' +      {1 \over 6} \int (A+x)'((A+x)\wedge  (A+x))'~.}
Prime defines an isomorphism
$$': \Omega^{(0,p)}(\wedge^q T_M) \longrightarrow \Omega^{(q,p)}~,$$
given by the contraction with holomorphic $3$-form $\Omega$.
Both $x$ and $A$ satisfy the constraint
$\partial A'=\partial x'=0$.
The variation of complex structure is given as
\eqn\var{\bar \partial \longrightarrow  \bar \partial+ \phi \cdot
\partial~,}
where $\phi \in H^{(0,1)}(T_M)$. Under this variation the holomorphic
$3$-form
varies according to
\eqn\omm{\Omega \rightarrow \Omega + \phi'+(\phi \wedge \phi)' +(\phi
\wedge
\phi \wedge \phi)'  }
In fact we will only need the linear term. In this discussion we will
assume
that we are making an analytic continuation  away from the geometric
slice
which allows us to relax the condition
$\partial+ \bar \partial = d$  fixed and treat $\partial$ and $ \bar
\partial$
independently. Under the variation
\var\ $\partial$ does not change. Let us postulate the following
transformation
law. We will ses in a moment
 that it is indeed a symmetry of KS action for  the field $Y=A+x$
\eqn\kstrans{Y'  \longrightarrow (Y+ \delta Y)^{\xi}=Y'-\phi'~,}
where ``$^{\xi}$" defines a deformed prime operation with respect to
the new
holomorphic $3$-form \omm. This transformation rule implies that
the variation
$\delta Y \in \Omega^{(0,1)}(T_M) \oplus \Omega^{(0,2)}(\wedge ^2
T_M)$.
The deformed massless mode $x+ \delta x$ is killed
by the new operator $\partial_{\rm new}=\bar \partial + \phi \cdot
\partial$.
Projecting on the kernel of $\partial_{\rm new}$  we recover
the transformations rules for massless and massive modes
\eqn\kssep{\eqalign{  (\delta x)'&=-\phi' - x \perp \phi' - {1 \over
\bar
\partial} \partial (\phi \wedge x)' \cr
 (\delta A)'&=   -A \perp \phi' + {1 \over \bar \partial} \partial
(\phi \wedge
x)'\cr }}
Operation $\perp$ defines a contraction of holomorphic vector indices
and
naturally
replaces $m(\cdot, \cdot)$.
The above formulas define a connection on the space of massless
(massive) modes
which should  be compared with \newfield a-\newfield b.
It is straightforward to check that \kssep\ is indeed a symmetry of
KS action
\eqn\kssymm{S[A +\delta A, x +\delta x;J+\delta J]=S[A, x ;J]- {1
\over 2}\int
x' (x \wedge \phi)'~.}
This relation is just an infinitesimal form of background
independence similar
to \ktsp.
The discussion of the previous section is fully applicable to KS
theory.
Equation \kssep\ defines a parallel transport  $x(j)$ (with boundary
condition
$x(J)=x$)
on the space of zero modes. The solution of equation
$x(j)=0$ determines a new complex structure $\tilde J$.  The global
form of
background
(in)depenedence presented in \bcov\  relates KS actions for  $J$ and
$\tilde
J$.

\newsec{Connection}

\subsec{Differential geometry of  K\"ahler forms.}

The first equation of  \newfield a\ defines a differential equation
on the
space ${\cal K}^{1,1}(M)$ of K\"ahler forms on $M$.  This is an
infinite-dimensional vector space.  In Appendix A we describe a
special
foliation\foot{Specifically,   ${\cal K}^{1,1}(M)$ is {\it foliated}
by
$b_{1,1}$-dimensional  (smooth) surfaces, called {\it leaves}. This
means  that
 every point $\omega \in {\cal K}^{1,1}(M)$ belongs to one and only
one such
surface. The leaves depend smoothly on the point  of  ${\cal
K}^{1,1}(M)$.}
${\cal F}$ of  ${\cal K}^{1,1}(M)$  (the Hodge foliation).  A tangent
space to
leaf of  ${\cal F}$ at  the point  $\omega \in {\cal K}^{1,1}(M)$
consists of
 $(1,1)$-forms  harmonic with respect to the K\"ahler structure
$\omega$.
Obviously, this means that the leaves are $b_{1,1}$-dimensional.
Locally, one
can introduce the coordinates
$\{z_1,\ldots,z_{b_{1,1}},a_1,a_2,\ldots\}$ on
${\cal K}^{1,1}(M)$  such that $\{z_1,\ldots,z_{b_{1,1}}\}$ are the
coordinates
along the leave and $\{a_1,a_2,\ldots\}$ parameterize different
leaves.

Over  ${\cal K}^{1,1}(M)$, one can consider a few  vector bundles.
One is  the
(infinite-dimensional) bundle $\cal V$ of massless modes with fibers
${\cal
H}_\omega$ consisting of solutions of equations  $dx=0$  and
$d^{c\dagger}x=0$.
The other is {\cal H} --- the bundle of  ``gauge-fixed''  massless
modes,
satisfying  an extra equation $d^{\dagger}x=0$. By definition, the
solutions
are the harmonic forms.  Restricting ourselves to $(1,1)$-forms, we
obtain a
bundle  {\cal H}$_{1,1}$ --- the  bundle of tangent directions to
leaves of the
Hodge foliation ${\cal F}$.
All this zoo has already appeared in our discussion.  The action
$S_0[x,\;
\omega]$ is defined as a  function on ${\cal V}$ (and the AKS action
$S[K,x;\,
\omega]$ essentially is a function on ${\cal V}\times \Omega ^3(M)$).
The
transformation \newfield a\ can be considered as a  differential
equation on
the section $x(\omega (z))$ of the bundle ${\cal T}L$
\eqn\eol{\partial_a x(\omega) + s(\partial_a(\omega) ,
x(\omega))=\partial
_a(\omega)}
 where ${\cal T}L$ is defined as  {\cal H}$_{1,1}$, restricted to a
leaf of
$\cal F$ and $\partial _a\equiv \partial /\partial z^a$ are the
coordinate
vectors (so that $\partial _a(\omega)$ are harmonic forms).  The
sections
$x(\omega)$ of  ${\cal T}L$ are vector fields on leaves, in
components
$x(\omega )=x^i(\omega )\partial_i$.

 In principle, we could continue  using the Hodge foliation  ${\cal
F}$ and the
bundles $\cal V$ and {\cal H}. In  particular, one can show that
$\partial_a +
s(\partial_a , \cdot)$ is a natural flat connection on ${\cal T}L$.
This
connection can be trivialized by choosing  {\it flat coordinates}
$\{t^i\}$
along the leaf.
Then  \eol\  defines the vector fields along the leaves, linear  in
$t^i$.

  For our purposes it seems more natural though to make use of  the
symmetries
to pass to more conventional finite-dimensional objects, as we do in
the next
section.  Still, it is convenient to keep in mind the picture just
described
since it explains clearly the geometric meaning of  $s(\delta \omega
,\cdot)$.

\subsec{Reduction to  a finite-dimensional picture.}

Although AKS is naturally defined  on the infinite-dimensional space
$\cal V$
of parameters $x$ and $\omega$  discussed above, one can effectively
reduce
$\cal V$ to a finite-dimensional object using the symmetries
established in the
Section 3.  To fix the gauge symmetry one requires $d^{\dagger}x=0$
which
reduces $\cal V$ to the bundle {\cal H} with finite-dimensional
fiber.  Next,
AKS is what we call in  Section 3 a K\"ahler topological theory:
essentially it depends only on the {\it cohomological class} of the
K\"ahler
form.

Let us consider  the transformation \newfield a\ : $x \rightarrow
\tilde x= x +
\delta \omega-s(\delta \omega,x) $.  In that equation, $x$ and
$\delta \omega$
are harmonic with respect to $\omega$ and  $\tilde x$ is harmonic
with respect
to $\omega +\delta \omega$. Now let us interpret \newfield a\ as
an equation on
the {\it cohomological class} of $x(\omega )$. To make sure this is a
consistent interpretation one should check that the class $[\tilde
x]$ depends
only on the class $[x]$. Indeed,  if $x\rightarrow x+d\alpha$ then
\eqn\cotran{\eqalign{
\tilde x&\rightarrow \tilde x +d\alpha -P_{{\rm Ker}\,d}m(d\alpha
,x)=d(\alpha
+m(\alpha ,x))-P_{{\rm Ker}\,d}d^{c\dagger}(\alpha x) \cr
&=\tilde x+d\Bigl(\alpha +m(\alpha ,x)+{d^{c\dagger}\over d}(P_{{\rm
Ker}\,d}\alpha x-h_{P_{{\rm Ker}\,d}\alpha x})\Bigr).
}}

 The same argument  shows that  $[\tilde x]$ depends only on the
class $[\delta
\omega ]$ of the variation of  the K\"ahler form.  This motivates one
to
consider a bundle $\cal C$ with fibers $H^*(M)$ over the {\it
K\"ahler cone}
${\rm K} \subset H^{1,1}(M)$. The bundle is defined by the connection
\eqn\connection{D_a =\partial_a + s(\partial_a(\omega), \cdot)~,}
where we introduced the coordinate system $\{z^i\}$ on K and
$\partial_i\equiv
\partial/\partial z^i$ (~so that
$\partial_i(\omega)\in~H^{1,1}(M)$~).
Then the ``cohomological version'' of equation \newfield a\
which can be written as\foot{Note that the equation \connection\  can
also be
written as ${\cal D}_a x=0$, where ${\cal D}_a x=D_a x +\partial
_a(\omega)$ is
the {\it affine} connection, associated with the linear connection
$D_a$.
(~${\cal D}_a$ is called the Cartan connection.)  This remark should
explain
what we ment in Section 2  describing the accessories of global
background
independence. }
\eqn\eoc{D_a x+ \partial_a(\omega)  =0}
defines a section $x(\omega)$ of  the tangent bundle to the K\"ahler
cone
${\cal T}{\rm K}\subset \cal C$.
The sections $x(\omega(z))$ of  ${\cal T}L$ are vector fields on K,
in
components $x(\omega(z) )=x^i(z )\partial_i$. The equation \eoc\
written in
components takes the following form:
\eqn\eoccom{\partial_a x^i + \Gamma ^i_{ab}(z)x^b=\delta ^i _a,}
where $\Gamma ^i_{ab}\partial _i(\omega )=s(\partial _a(\omega
),\partial
_b(\omega ))$.

 We will demonstrate that the connection $D_a$
is flat which implies that  the system \eoc\  is integrable.
It is convenient to present  \connection\ as
\eqn\decomp{
D_a=\partial _a +[\Lambda ,\partial _a (\omega )]-\Lambda (\partial
_a (\omega
))=\partial _a +[\Lambda ,\partial _a (\omega )]-\partial _a
\log{{\rm
Vol}_\omega .}}
The last term can be interpreted as a connection on the  flat line
bundle over
the K\"ahler cone.  It suffices to prove that $\nabla _a$ is
flat\foot{It is
worth mentioning that the operator  $\nabla _a=\partial _a +[\Lambda
,\partial
_a (\omega )]$
is a connection preserving the intersection form while $D_a$
preserves the
intersection form normalized to a unit  of  volume.
}.
\eqn\flawl{\eqalign{
[\nabla _a,\nabla _b]=[\partial _a \Lambda ,\partial _b (\omega )
]-[\partial
_b \Lambda ,\partial _a (\omega )]+\bigl[[ \Lambda ,\partial _a
(\omega )], [
\Lambda ,\partial _b (\omega )] \bigr].
}}
Let us compute the first two terms in \flawl. Since   $\partial _a
\Lambda
={1\over 2}[\Lambda ,[\Lambda , \partial _a(\omega )]],$
\eqn\Vars{\eqalign{
[\partial _a \Lambda ,\partial _b (\omega )]&={1\over 2}[[\Lambda
,\partial
_b(\omega )],[\Lambda , \partial _a(\omega )]]+{1\over 2}[\Lambda
,[[\Lambda
,\partial _b(\omega )], \partial _a(\omega )]] \cr
}}
The first summand in \Vars\ is antisymmetric in $a,\,b$ and the
second one is
symmetric. Substituting back to \flawl\ one finally obtains
\eqn\flawll{\eqalign{
[\nabla _a,\nabla _b]=\bigl[[ \Lambda ,\partial _b(\omega )], [
\Lambda
,\partial _a (\omega ) ] \bigr]+\bigl[[ \Lambda ,\partial _a (\omega
) ], [
\Lambda ,\partial _b (\omega ) ] \bigr]=0.
}}

Since the connection $D_a$  is flat there are sections
$\{e_{\alpha}\}$ that
trivialize the bundle: each section can be expressed as linear
combination of
$\{e_{\alpha}\}$.  In particular any solution of  $D_a   \xi=0$ is a
linear
combination of
$\{e_{\alpha}\}$ with constant coefficients.  The tangent bundle to
the
K\"ahler cone can be identified with the
subbundle of  $(1,1)$-forms of ${\cal C}$.  Let us consider the
subset
$\{e_i\}$ of $\{e_\alpha \}$ that generates  $H^{(1,1)}$ in each
fiber.  The
sections  $e_i$ are the vector fields on K. As a consequence of
flatness of
$D_i$ all the $e_i$ commute with each other.  Therefore there exists
a
coordinate system $\{t^i\}$ on a K\"ahler cone such that  these
vector fields
are tangent to the coordinate lines: $e_i=\partial /\partial t^i$. We
call
$\{t^i\}$ the flat coordinates.

 It is  instructive to find the flat coordinate $t$ in the simplest
case when
dim K=1.
One-dimensional  K\"ahler cone K  can parameterize the linear
coordinate
$\omega =y\, \sigma$, where $\sigma$ is any fixed $(1,1)$-form.
Computing  the
AKS connection \connection\ one  finds
$D_y=\partial _y+2y^{-1}$.   Solving $D_y e=0$ one obtains
$e=y^2\partial _y$
so that $t=y^{-1}$.

Using the flat coordinates one can immediately write down the
solution of
\eoc. Namely,

\vskip3mm
\eqnn\linsec
\hskip7pc \boxit{\vbox {\hsize 11pc $$x(t)=x^i  {\partial
\over\partial
t^i}=\big(t^i-t_0^i\big) {\partial \over\partial t^i}~,$$}}\hskip6pc
\raise1pc
\hbox{\linsec}

\noindent
where $B^i$ are the constants fixed by the initial data. Now one can
see that
for any initial data there is  a unique point  $[\tilde \omega
]=(t_0^1,\ldots,t_0^n)$ on the K\"ahler cone K  where the solution
$x$
vanishes.

In the previous section we used the slightly  different statement
that
for any
$x$ and any K\"ahler  {\it form}  $\omega$
one can find a K\"ahler {\it form} $\tilde \omega$ for which the
classical
solution $K_0$ vanishes. Now we can explain this.
The classical solution $K_0$ depends on the K\"ahler form $\omega$
and is
uniquely determined by cohomological class $[x]$ of $x$. Let us find
the
solution  \linsec\  of \eoc\ for the initial (cohomological) data
$[\omega]$
and $[x]$.
This solution can be promoted to the solution of \eol\ on the
particular leaf
$L_ \omega$ of the Hodge foliation specified by   $\omega \in
L_\omega$. There
is a one-to-one correspondence between the points on the leaf and the
points on
the K\"ahler cone. Then at the point $\tilde \omega \in L_\omega$
corresponding
to $[\tilde \omega ]\in \,$K the harmonic representative of
$[x(\tilde \omega
)]$ vanishes and so does $K_0[x(\tilde \omega )]$.

\def\Boxit#1{\vbox{\hrule\hbox{\vrule  \kern3pt
\vbox{\kern-15pt #1\kern-7pt}\kern3pt\vrule}\hrule}}

\newsec{Hamiltonian approach to AKS.}
\subsec{Canonical variables, Hamiltonian and constraints.}
In this section we return to the full AKS theory
 \fact\
described in Section 3.
We remind the reader that the ``superparticle field"  is defined as a
linear
combination ${\cal K}=\sum_{q=0}^{2n-1} K_q$,  where $K_q \in  \Omega
^{q}
(M)$.  The components with degrees $0\leq q(K) \leq n-1$ are called
fields,
while the  components  with degrees  $n\leq q(K) \leq 2n-1$ are
called
antifields.    The field $\cal K$  satisfies\foot{Thus $\cal K$ is
massive. The
massless mode is denoted by $x=\sum x_q$. } ${\cal
K}=d^{c\dagger}{\cal Z}$, or
in components  $K_q=d^{c\dagger}Z_{q+1}$.  Note that there is no
dynamical
field of  degree $2n$.
The action written in components is
$${1 \over 2 g^2}\int \sum_{p+q=2n-2} K_p {1 \over d^{c\dagger}}
dK_{q}
+  {1
\over 6 g^2}\int \sum_{p+q+r=2n}(x_p+K_p) \wedge (x_q+K_q) \wedge
(x_r+K_r)~.$$
It is important that the top component $K_{2n-1}$ does not have any
kinetic
term. Thus it is not dynamical.

We will  consider the Hamiltonian formulation of  AKS theory. It is
not
covariant. Even worse, splitting off of 1-dimensional time spoils
complex
geometry.
 Still this is the safest way to introduce the path integral.
Besides, in the
Hamiltonian approach  the basic physics of the model appears the most
clearly.
Also we will be able to use the wisdom accumulated in 3-dimensional
Chern-Simons theory \ref\wicsI{E.~Witten, Commun. Math. Phys. {\bf
121}  (1989)
351},
\ref\italcs{S.  Axelrod, UMI-91-24878-mc, Ph.D.Thesis}.

To begin, one should identify the time coordinate.  We will do this
in a way
which is not  quite general but has an advantage of preserving as
much of
complex geometry as possible. Assume that the manifold $M$ has a
structure of a
direct product  $M=S\times {\bf T}^2$ where $S$ is $n-1$-dimensional
complex
K\"ahler  and ${\bf T}^2$ is a 1-dimensional complex torus.  As a
real
manifold, ${\bf T}^2={\bf T}_\sigma^1\times {\bf T}_t^1$.  We call
the time the
coordinate $t$ parameterizing the circle ${\bf T}_t^1$. Then the
``space'' is
$S\times {\bf T}_\sigma^1$.  Also, let us choose a special  K\"ahler
structure
on $M=S\times {\bf T}^2$  which is $\omega =\omega _S\otimes 1+
1\otimes
(dt\wedge d\sigma )$, where $\omega _S$ is a K\"ahler structure on
$S$. Once we
know the physics for this particular $\omega$, we can move along
K\"ahler cone
of $M$ using the methods described  above.

The relation $K_{q-1}=d^{c\dagger}Z_{q}$ now reads as
$K_{q-1}=d_S^{c\dagger}Z_{q}+i(\partial _\sigma )\dot{Z}_{q}-\partial
_\sigma
i(\partial _t )Z_{q}$, where $d_S^{c\dagger}$ is a differential
operator on
$S$,  dot stands for differentiation  with respect to time and
$i(\partial _t
)$,  $i(\partial _\sigma )$ denote the contractions with the
coordinate frame
vectors.  This relation contains a time derivative.   The simplest
way
to take it
into account is to  write the  action in terms of $Z_q$.
\eqn\bez{\eqalign{ S[Z_p]=&{1 \over 2 g^2}\int \sum_{p+q=2n}
d^{c\dagger}Z_{p}\wedge dZ_{q}   +\cr
&{1 \over
6 g^2}\int \sum_{p+q+r=2n}(x_p+d^{c\dagger}Z_{p+1}) \wedge
(x_q+d^{c\dagger}Z_{q+1}) \wedge (x_r+d^{c\dagger}Z_{r+1})~.\cr}}
Let us decompose $\cal Z$ as  ${\cal Z}={\cal Z}^0+{\cal Z}^1\wedge
d\sigma$,
where ${\cal Z}^1=i(\partial _\sigma){\cal Z}$. Defined as a
functional of
$Z_q$ and $\partial _\mu Z_q $ the action depends on the velocities
$(\dot{Z}^0_q)_{\mu _1\dots \mu_{q-1}}$  linearly.  Indeed,  the term
$d^{c\dagger}{\cal Z}=d_S^{c\dagger}{\cal Z}+\dot{\cal Z}^1-\partial
_\sigma
i(\partial _t) {\cal Z}$ cannot produce such velocity at all. The
term  $d{\cal
Z}$ produces the combination $\dot{\cal Z}^0\wedge dt $ which has to
be
multiplied by $(d_S^{c\dagger}{\cal Z}^1 - \partial _\sigma
i(\partial _t)
{\cal Z}^1)\wedge d\sigma$.  Hence the  momenta $ {\partial {\cal
L}\over
\partial \dot{Z}^0_q}$ corresponding to ${\cal Z}^0$ can be written
in terms of
 the spacial derivatives of ${\cal Z}^1$ only. The Hamiltonian   $H=
{\partial
{\cal L}\over \partial \dot{Z}_q}\dot{Z}_q-L$ is independent of these
momenta.
We conclude that  ${\cal Z}^0$ is conserved  and serves  just as a
parameter\foot{We discuss the corresponding secondary constraint
below.}. This
is a consequence of the obvious symmetry ${\cal
Z}\longrightarrow{\cal
Z}+d_S^{c\dagger}{\cal W}$.

Also, let us give a closer look to the equations of motion. One can
easily see
that  no second time derivatives $i(\partial _t)\ddot{\cal Z}$ of
the {\it
temporal} components  can be found.  Therefore, $i(\partial _t){\cal
Z}$ is not
dynamical in a usual sense. This happens because of the gauge
symmetry.
To proceed, we should  choose the gauge fixing.  From the above
discussion it
follows one can consistently take the {\it temporal gauge}
$i(\partial
_t){\cal Z}$=0. Geometrically it means that  $\cal Z$  is a
differential form on
 $S\times {\bf T}_\sigma^1$ depending on $t$ as a parameter
\foot{Clearly, this gauge  does not  exist for the top component
$Z_{2n}$: it
cannot be made into a form on $S$. But since the field
$K_{2n-1}=d^{c\dagger}Z_{2n}$ is not dynamical anyway, we can simply
forget
about  $Z_{2n}$.}.
We should also fix the massless modes $x_q$ satisfying
$dx_q=d^{c\dagger}x_q=0$.   As usual we take the  harmonic
representatives on
$M$.  Obviously, they can be decomposed as
$x_q=x^{00}_q+x^{10}_q\wedge dt +x^{01}_q\wedge d\sigma
+x^{11}_q\wedge dt
\wedge d\sigma$ where $x_q^{nm}$ are the harmonic forms on $S$. For
the sake of
simplicity, let us assume that  $i(\partial)x_q=0\Longleftrightarrow
x_q^{10}=x_q^{11}=0$.  In fact, imposing these constraints  we loose
a part of
the information about  the topology of the space-time $M$ in $t$
direction.
In
particular, the Wilson lines   along ${\bf T}^1_t$ are excluded.

There is  a natural  Poisson bracket on differential forms  on
$S\times {\bf
T}_\sigma^1$ which may formally be written as
\eqn\pb{\{ K_q (z), Z_{p}  (w) \}_ {_{\rm P.B.}} =
 \delta_{p+q,2n-1}\, \omega _S^{n-1}(z)\wedge d\sigma \,\delta(z,w)~.
}
 This bracket paires $\Omega ^p(S\times {\bf T}_\sigma^1)$ with
$\Omega
^{2n-1-p}(S\times {\bf T}_\sigma^1)$.
As usual, \pb\ means that  the Poisson bracket of two functionals is
given by
\eqn\fbrac{\lbrace{ F,L \rbrace}_ {_{\rm P.B.}}=
\int  \sum_{p+q=2n-1}    {\delta F \over \delta K_q}
 {\delta L \over \delta Z_p} -
{\delta F \over \delta Z_p} {\delta L \over \delta K_q} ~.}

Let us write down the action
 \fact\
in the temporal gauge. Since $\cal Z$ and $x$ are differential forms
on
$S\times {\bf T}_\sigma^1$, so is ${\cal K}=d_S^{c\dagger}{\cal
Z}+i(\partial
_\sigma )\dot{\cal Z}$.
The cubic term  $({\cal K}+x)^3$ equals zero as a differential form
of degree
$2n$ on $2n-1$-dimensional manifold $S\times {\bf T}_\sigma^1$. Using
the above
decomposition $Z_q=Z^0_q +Z^1_q \wedge d\sigma $ one can write the
action as:
\eqn\gaufixa{S[{\cal Z}]={1 \over 2 g^2}\int dt  \int_{{\bf
T}_\sigma^1}d\sigma
\int_S \sum_{p+q=2n} \dot{Z}^1_p\wedge\dot{Z}^1_q+2\dot{Z}^1_p\wedge
d_S^{c\dagger}{ Z}_q^0 ~}
(we used integration by parts.)
The dynamical variables are $Z_p^1$ and their canonical momenta
$K^0_{2n-1-p}=d_S^{c\dagger}{ Z}_{2n-p}^0+\dot{Z}_{2n-p}^1$ --- the
restrictions of  $K_p$ to $S$.  The canonical momenta for $Z^0_p$ are
zero. The
constraint   $d_S^{c\dagger}K^0_{q-1} =d_S^{c\dagger} \dot{Z}^1_q=0$
and the
equation of motion $d_S^{c\dagger}\dot{Z} _q^0+\ddot{Z}_q^1=0$
together are
equivalent to  $\partial_t{\cal K}=0$.  The Hamiltonian
 \eqn\hamilt{H[{\cal K}^0, {\cal Z}^1]={1\over 2 g^2}\int _{{\bf
T}_\sigma^1}d\sigma \int_S {\cal K}^0\wedge {\cal K}^0 - 2{\cal
Z}^0\wedge
d^{c\dagger}_S{\cal K}^0~}
is independent\foot{This is no longer true if  $i(\partial _t)x\neq
0$. In
general for $x=y+z\wedge dt$ the equation of motion $\partial_t{\cal
K}+d^{c\dagger}_S(y\wedge z+z\wedge {\cal K})=0$ shows that  $H[{\cal
K}^0,
{\cal Z}^1]$ is a functional of both canonical coordinates.} of
${\cal Z}^1$.

But  this is not the whole story yet.  Gauge fixing $i(\partial
_t){\cal Z}=0$
produces a bunch of  secondary constraints.  These constraints can
easily be
found  if we notice that together with the dynamical equation
$\partial_t{\cal
K}=0$ they should reproduce the AKS equations of motion
$d{\cal K}+d^{c\dagger}(({\cal K}+x)^2)=0$ in the temporal gauge
$i(\partial
_t){\cal K}=0$.
Taking the decomposition ${\cal K}={\cal K}^0+{\cal K}^1\wedge
d\sigma$ where
${\cal K}^1=i(\partial _\sigma){\cal K}$ and the similar
decomposition for the
massless modes one finds
\eqn\dinconstr{\eqalign{
{\cal C}_1=d_S{\cal K}^0+{1\over 2}d_S^{c\dagger}(({\cal
K}^0+x^0))^2&=0 \cr
{\cal C}_0=\partial _\sigma {\cal K}^0+d_S{\cal
K}^1+d_S^{c\dagger}(({\cal
K}^0+x^0)\wedge ({\cal K}^1+x^1))&=0 \cr
}}
Together with  ${\cal K}^1=d_S^{c\dagger}{\cal Z}^1$, the fact that
${\cal
K}^0$ is the canonical conjugate for  ${\cal Z}^1$ and
$d_S^{c\dagger}{\cal
K}^0=0$ the relations \dinconstr\  produce all the  constraints\foot{
The
background fields $x^0$ and $x^1$ are independent of $\sigma$. For
every
$\sigma$, we can solve the first equation in \dinconstr\ which is the
equation
of motion of  AKS theory {\it  living on} $S$. Given $x^0$, the
solution is
unique  modulo  gauge transformations. This means that as we move
along ${\bf
T}_\sigma ^1$,  the field ${\cal K}^0$ can only change by the gauge
transformation. The second equation in  \dinconstr\  tells us this is
indeed
so: the infinitesimal shift along ${\bf T}_\sigma ^1$ is equivalent
to
the gauge
transformation with parameter ${\cal K}^1+x^1$.}.

Both fields ${\cal K}^0$ and ${\cal Z}^1$ are defined over  the
product
$S\times {\bf T}_ \sigma              ^1$.  There is a remaining
gauge
symmetry\foot{From now on, there will appear the whole zoo of gauge
groups.
Reader may find it convenient to have a glossary. By ${\cal G}_M$ we
denote the
original  gauge group \gauge. Similarly, ${\cal G}_S$ denote the
gauge group of
AKS theory on the manifold $S$. Its parameters are the differential
forms on
$S$. We also need a group  $\tilde{\cal G}_S$ which is ${\cal G}_S$
{\it with
$\sigma$-- dependent  parameters $\Omega ^*(S)\otimes \Omega ^0({\bf
T}^1\sigma
)$}.  The latter has a central extension $\hat{\cal G}_S$   to be
described
below.   The subgroup of  ${\cal G}_M$ preserving the temporal gauge
is called
$\tilde{\cal G}_R$. Its subgroup  ${\cal G}_R \subset \tilde{\cal
G}_R$  is
obtained by restricting to $\sigma$--independent parameters. One has
${\cal
G}_M\supset \tilde{\cal G}_R \supset \tilde{\cal G}_S \supset {\cal
G}_S$. } $
\tilde{\cal G}_R$ --- a subgroup of the AKS gauge group preserving
the temporal
gauge.  The gauge parameter of  $ \tilde{\cal G}_R$ should satisfy
$\dot{\alpha}=0$ and $i(\partial _t)\alpha=0$. It can be decomposed
as
$\alpha=\alpha ^0 +\alpha ^1\wedge d\sigma$, both $\alpha ^0$ and
$\alpha ^1$
being annihilated by $d_S^{c\dagger}$. Using the explicit
parameterization of
${\rm Ker}\,d_S^{c\dagger}$  introduced in Section 3, one can write
$\alpha
^i=d_S^{c\dagger}\chi ^i +h^i$ where $h^i$ are harmonic. The gauge
transformations\foot{It is easy to see that actually $\tilde{\cal
G}_R \cong
\tilde{\cal G}_S\triangleright \Lambda ^*(R_{\tilde{\cal G}_S})$ ---
a
semidirect product of  $ \tilde{\cal G}_S$ with the external algebra
of  its
adjoint  representation $R_{\tilde{\cal G}_S} $.} $ \tilde{\cal G}_R$
written
in terms of  $\chi ^i$, $h^i$ are:
\eqn\regasyI{\eqalign{
\delta _{0} {\cal K}^0 &= d_Sd_S^{c\dagger}\chi ^0 +  d_S^{c\dagger}
((d_S^{c\dagger}\chi ^0+h^0) \wedge ( {\cal K}^0+x^0))\cr
\delta _0 {\cal Z}^1&= \partial _\sigma \chi ^0+ (d_S^{c\dagger}\chi
^0+h^0)
\wedge ( d_S^{c\dagger}{\cal Z}^1+x^1)\cr
}}
and
\eqn\regasyII{\eqalign{
\delta _1 {\cal K}^0 &=0\cr
\delta _1 {\cal Z}^1&= d_S\chi ^1 +  (d_S^{c\dagger}\chi ^1+h^1)
\wedge ( {\cal
K}^0+x^0)~.\cr
}}

The remaining gauge symmetry can be fixed. To fix \regasyI\ we impose
$d_S^{\dagger}{\cal K}^0=0$ and $\partial _\sigma{\cal Z}^1=0$. To
fix
\regasyII\ we add  $d_S^{\dagger}{\cal Z}^1=0$.  Then the constraints
\dinconstr\ can be solved to find  ${\cal K}^0$ and ${\cal K}^1$ in
terms of
the massless modes $x^0$ and $x^1$ using the series
 \Sol.
So defined,  ${\cal K}^0$ and  ${\cal K}^1$ are independent of
$\sigma$.  The
space of solutions can be identified with the double ${\cal
H}_S\oplus {\cal
H}_Sd\sigma$  of the space ${\cal H}_S$ of harmonic forms on $S$.

We expect that if we did not set the temporal components
$x^{10}=x^{11}=0$, we
would obtain the full space ${\cal H}_M \sim {\cal H}_S\oplus {\cal
H}_Sd\sigma
\oplus{\cal H}_Sdt \oplus {\cal H}_Sd\sigma dt$ of harmonic forms on
$M$. In
fact, we have already established  this in the section about the BV
formalism.
We see that  AKS turns out to be a finite-dimensional system.  Its
dynamics is
governed by the nonzero Hamiltonian.
This is to be compared with 3-dimensional Chern-Simons theory which
is also a
finite-dimensional system but with a zero Hamiltonian.

\subsec{Classical and quantum symplectic reduction.}
The gauge transformations \regasyI --\regasyII\ with $\alpha
^0=d^{c\dagger}_S
\chi ^0$ and $\alpha ^1=d^{c\dagger}_S \chi ^1$ generate a (normal)
subgroup
$\tilde{\cal G}_R^0 \subset \tilde{\cal G}_R $. The action of this
subgroup
is (almost) Poisson.  Indeed,  the constraints \dinconstr\  generate
the flows:
\eqn\poigas{\oint d\sigma \int_S \chi ^i(z)\wedge \{{\cal
C}_i,\,F[{\cal
K}^0,{\cal Z}^1] \}_ {_{\rm P.B.}}=\delta _i\, F[{\cal K}^0,{\cal
Z}^1]~.}
We   compute the P.B.'s~between the constraints to find
\eqn\cexga{\eqalign{
\Big\{\int \xi ^0(x)\wedge {\cal C}_0,\,\int \zeta ^0(y)\wedge {\cal
C}_0\Big\}_ {_{\rm P.B.}}&=\int (d^{c\dagger}_S\zeta \wedge
d^{c\dagger}_S\xi
)\wedge {\cal C}_0 +\int \partial_\sigma \xi ^0 \wedge
d_Sd^{c\dagger}_S\zeta
^0\cr
\Big\{\int \xi ^0(x)\wedge {\cal C}_0,\,\int \zeta ^1(y)\wedge {\cal
C}_1\Big\}_ {_{\rm P.B.}}&=\int (d^{c\dagger}_S\zeta \wedge
d^{c\dagger}_S\xi
)\wedge {\cal C}_1  \cr
\Big\{\int \xi ^1(x)\wedge {\cal C}_1,\,\int \zeta ^1(y)\wedge {\cal
C}_1\Big\}_ {_{\rm P.B.}}&=0  \cr
}}
The relations \cexga\ show that  \poigas\ furnish a representation of
{\it
central extension} $\hat{\cal G}_R^0$ of the
gauge group  $ \tilde{\cal G}_R^0$. A cocycle in the right hand side
of \cexga\
appears due to nontrivial $\sigma$ dependence of the gauge
parameters.

We would like to leave an interesting object $\hat{\cal G}_R^0$ for
further
investigation.  Our immediate aim is to obtain  the physical phase
space. From
the above discussion  it follows that  in a sense,
$\sigma$--dependence is the
pure gauge.  This motivates one to consider a restriction to
$\sigma$--independent fields and the action  of \regasyI --\regasyII\
with
$\sigma$--independent gauge parameters. (Essentially this returns us
to AKS
theory considered, however,  on the manifold $S$.)

So let us consider a phase space ${\cal V}=\{{\cal K}^0,{\cal
Z}^1\,|\,{\cal
K}^0\in {\rm Ker}\,d^{c\dagger}_S,\,{\cal Z}^1\in \Omega ^*(S)/{\rm
Ker}\,d^{c\dagger}_S\}$ with Poisson bracket  given by
$$\{ {\cal K}^0(z), {\cal Z}^1 (w) \}_ {_{\rm P.B.}} =
 \omega _S^{n-1}(z) \,\delta(z,w)~. $$
 The  transformations \poigas\  with  $\sigma$--independent
parameters $\chi
^i$ act on functions on $\cal V$ furnishing a representation of the
gauge group
${\cal G}_R^0$.  In particular, these transformations preserve the
ideal  of
the set $({\cal C}_1=0,\,{\cal C}_0=0)$.

(Indeed, doing the hamiltonian reduction, first we take the subspace
of  $\cal
V$ where ${\cal C}_1={\cal C}_0=0$.  Then we impose the secondary
constraints
$\{{\cal C}_1,\,\cdot \}=\{{\cal C}_0,\,\cdot \}=0$.  As \poigas\
shows, this
is equivalent to taking  ${\cal G}_R$-invariants.
As a result of the reduction, we obtain  a single point, since the
massless
fields $x^0$ and $x^1$ are kept fixed all the time. )

Now we can turn to quantization.  First  we should choose a
polarization on the
phase space $\cal V$. It is convenient to work in   ``${\cal
K}^0$--representation'':   wave functions $\Psi [{\cal K}^0]$ are
functionals
of  ${\cal K}^0$;  the canonical  conjugate of  ${\cal K}^0$  which
is ${\cal
Z}^1=i\,\delta /\delta {\cal K}^0$ acts on  $\Psi [{\cal K}^0]$ by
differentiation. Wherever  the ordering problem occurs, we use the
``$qp$''
prescription -- put the momenta to the right.

The gauge symmetry \regasyI --\regasyII\ is realized by the
differential
operators acting on wave functions. Obviously, the generator $\delta
_1$   acts
by multiplication by  ${\cal C}_1=d_S{\cal K}^0+{1\over
2}d_S^{c\dagger}(({\cal
K}^0+x^0))^2$. The generator $\delta _0$ is represented by the
differential
operator
\eqn\gegat{\hat{G}=d_S^{c\dagger}d_S\,i\,{\delta \over\delta {\cal
K}^0}+d_S^{c\dagger}\Big[({\cal K}^0+x^0)\wedge
(d_S^{c\dagger}\,i\,{\delta
\over\delta {\cal K}^0}+x^1)\Big]}

The first constraint \dinconstr\  means that  we should consider only
the wave
functions with support on the   space $ {\cal N}\equiv \{{\cal
K}^0\,|\,
d_S^{c\dagger}{\cal K}^0=0;\, {\cal C}_1=0\}$ ($\cal N$ consists of
solutions
of  AKS equations of motion on $S$). The gauge group ${\cal G}_S^0$
acts on
$\Omega ({\cal N})$ by $\hat{G}$. Then the second constraint
\dinconstr\  says
that the physical wave  functions satisfy\eqn\ginwf{\hat{G}\,\Psi
[{\cal
K}^0]=0.}
In other words, the physical wave function is gauge invariant\foot{If
we do not
restrict  to $\sigma$-independent fields, we  obtain $\hat{G}\Psi
[{\cal
K}^0]=(\partial_\sigma {\cal K}^0)\Psi [{\cal K}^0]$.  This means
that  $\Psi
[{\cal K}^0]$ furnishes a {\it projective} 1-dimensional
representation of the
gauge group: the gauge transformation changes only a phase of the
wavefunction.
Note that  the nontrivial phase variation is a consequence of
presence of
the parameter $\sigma$. This type of phenomena is usually referred to
as a
Berry phase. Equivalently, one can say that   $\Psi [{\cal K}^0]$ is
invariant
with respect to action of the central extension of the gauge group.}.
Together, two relations  \dinconstr\ describe the Hamiltonian
reduction of
$\Omega ^0({\cal V})$ with respect to  $\hat{\cal G}_R^0$.

Similarly to CS theory, one can write the solution to \ginwf\ in
terms of the
functional integral.
Consider a functional\foot{The functional integral should be computed
over
$\Omega ^*(S)/{\rm Ker}\,d^{c\dagger}_S$}
\eqn\pinwf{
\Psi [{\cal K}^0]=e^{i\int({\cal K}^0+x^0)\wedge x^1} \,\epsilon
_{\cal
N}\big[{\cal K}^0\big]  \int  {\cal D} {\cal Z}\, e\,^{i{1 \over g^2}
\int
d_S^{c\dagger}{\cal Z} \wedge d_S{\cal Z}+d_S^{c\dagger}{\cal
Z}\wedge ({\cal
K}^0+x^0)\wedge d_S^{c\dagger}{\cal Z} }}
where  $\epsilon _{\cal N}\big[{\cal K}^0\big]$ is a characteristic
function of
the set $\cal N$ and
\eqn\noname{\eqalign{&\int  {\cal D} {\cal Z}\, e\,^{i{1 \over g^2}
\int d_S^{c\dagger}{\cal Z} \wedge
d_S{\cal
Z}+d_S^{c\dagger}{\cal Z}\wedge ({\cal K}^0+x^0)\wedge
d_S^{c\dagger}{\cal Z}
}=  \cr
&\Big[ \int d\Omega \,{\rm Ber}\,'\, (d_S
d_S^{c\dagger}+d_S^{c\dagger}(
({\cal K}^0+x^0)\wedge d_S^{c\dagger}))\Big]^{1/2}~.\cr}}
Here ${\rm Ber}\,'$ denotes the Berezinian computed for nonzero modes
and
$d\Omega$ is a supermeasure on the space of zero modes of  the
operator  $L=d_S
d_S^{c\dagger}+d_S^{c\dagger}( ({\cal K}^0+x^0)\wedge
d_S^{c\dagger})$.
Obviously, \pinwf\ has a support on $\cal N$.  It is easy to see that
$\hat{G}\,\epsilon _{\cal N}\big[{\cal K}^0\big]=0$ (this should
already be
clear from \poigas).  Using ``the equations of motion'' for the
functional in
the exponent of the integral representation for the Berezinian one
can see that
 \ginwf\ is indeed satisfied.

The scalar product of two wave functions is defined as
$$\langle \Psi _1|\Psi _2\rangle =\int {\cal D} {\cal
K}^0\,\delta\big[{\cal
C}_1[{\cal K}^0]\big] \delta \big[d_S^{c\dagger}{\cal K}^0\big]\,
\overline{\Psi}_1[{\cal K}^0]\Psi _2[{\cal K}^0]$$
where the $\delta$--functionals can also be written as
 $$\delta\big[{\cal C}_1[{\cal K}^0]\big] \delta
\big[d_S^{c\dagger}{\cal
K}^0\big]=\int {\cal D}\lambda\,{\cal D}\mu\, e^{i\int {1\over
2}({\cal
K}^0+x^0)^2\wedge d_S^{c\dagger}\mu+{\cal K}^0\wedge
(d_S^{c\dagger}\lambda
-d_S\mu)}.$$

The gauge invariant measure on the configuration space which appears
in the
scalar product can also be interpreted as the invariant measure on
the gauge
group ${\cal G}_S$.  Indeed, the set $\cal N$ consists of  the
solutions
${\kappa}^0$ of the AKS equations of motion. For a given $x^0$, they
constitute
the orbit of that gauge group.  Introducing the coordinates
${\kappa}^0$ along
${\cal N}$ and the transversal coordinates $ \eta$ we can write
\eqn\meI{\eqalign{
\delta\big[{\cal C}_1[{\cal K}^0]\big] \delta \big[d_S^{c\dagger}\eta
\big]{\cal D} \kappa ^0 {\cal D}\,\eta &=\epsilon _{\cal N}\big[{\cal
K}^0\big]\Bigl[{\rm Ber}\,' \Bigr( {\delta {\cal C}_1\over \delta
\eta}\Bigl){\rm Ber}\,'(d_S^{c\dagger})\Bigr]^{-1}\,{\cal D} \kappa
^0\cr
&=\epsilon _{\cal N}\big[{\cal K}^0\big]\big[{\rm Ber}\,' (d_S +
\kappa
^0\wedge d_S^{c\dagger})\big]^{-1}\bigl[{\rm
Ber}\,'(d_S^{c\dagger})\bigr]^{-1}\,{\cal D} \kappa ^0~.
}}

On the other hand,  using  the gauge parameter $\alpha ^0$ of
\regasyI\ as a
coordinate along $\cal N$, we find
${\cal D} \kappa ^0= \Bigl[{\rm Ber}\,' (d_S + \kappa ^0\wedge
d_S^{c\dagger})\Bigr]\,{\cal D} \alpha ^0, $
so finally the measure\foot{To save the space we omit  the factor
$\epsilon
_{\cal N}\big[{\cal K}^0\big]\,$.} is just $\bigl[{\rm
Ber}\,'(d_S^{c\dagger})\bigr]^{-1}\,{\cal D} \alpha ^0$. Note that
this is
indeed a natural measure on the gauge group. Since the gauge
parameter should
satisfy the constraint  $d_S^{c\dagger}\alpha ^0=0$, we can write the
measure
as $\delta \bigl[d_S^{c\dagger}\alpha \bigr]\,{\cal D}\alpha $ and
extend
integration over the whole ${\Omega ^*(S)}$.  Integrating out the
transversal
coordinates, one recovers the factor   $\bigl[{\rm
Ber}\,'(d_S^{c\dagger})\bigr]^{-1}$.

\subsec{Path integral for AKS}
The measure for the Hamiltonian path integral in temporal gauge is
determined
by the Poisson bracket \pb. It can be written as
$${\cal D}\mu=[{\cal D}K_{2n-1}]\prod_{p=0}^{2n-2}[{\cal
D}K_p^0][{\cal
D}Z_{2n-1-p}^1]\,\delta[{\cal C}_0]\,\delta[{\cal
C}_1]\,\delta[d_S^{c\dagger}{\cal K}^0]~.$$
All the fields here are $\sigma$--dependent.  As in the relatively
simpler case
just considered, our aim is to reduce ${\cal D}\mu$ to the measure on
the
gauge group $\tilde{\cal G}_R$ generated by $\delta _0$ and $\delta
_1$.  By
$\kappa ^0$ and $z^1$ we denote the solutions of the constraints.
These
solutions are parameterized by $\alpha ^0$ and $\alpha ^1$ of
\regasyI--\regasyII. The transversal coodinates are $ \eta$ and $
\xi$.
Following the same steps as above, one finds:
\eqn\meII{\eqalign{
{\cal D}\mu &=\Bigl[{\delta ({\cal C}_1,  {\cal C}_0)\over \delta
(\eta , \xi
)}\Bigr]_{\rm Ber}^{-1} \bigl[{\rm
Ber}\,'d_S^{c\dagger}\bigr]^{-1}\,\Bigl[{\delta (\kappa ^0,
z^1)\over \delta (
\alpha^0 ,  \alpha ^1 )}\Bigr]_{\rm Ber} {\cal D} \alpha ^0\,  {\cal
D} \alpha
^1 \cr
&= \bigl[{\rm Ber}\,'d_S^{c\dagger}\bigr]^{-2}\,  {\cal D} \alpha
^0\,  {\cal
D} \alpha ^1
}}
which is a natural measure on the gauge group.

Since the action is gauge invariant,  we can compute the path
integral of  any
gauge invariant observable $\cal O$ to obtain the ``localization
formula'':
\vskip2mm
\eqnn\localiz
\hskip0pc \boxit{\vbox{\hsize23pc $${\langle {\cal O}\rangle =\int
{\cal
D}\mu\, {\cal O}[{\cal K}^0, {\cal Z}^1]\, e^{-S[{\cal K}^0,{\cal
Z}^1]}={\rm
Vol}( {\cal G})\, {\cal O}[\kappa ^0, z^1]\, e^{-S[\kappa ^0,z^1]}
}$$}}\hskip1.2pc \raise1pc \hbox{\localiz}
\vskip2mm

 Let us briefly discuss the general situation  when $i(\partial _t)x
\neq 0$.
For $x=y+w\wedge dt$ the equation of motion $\partial_t{\cal
K}=-d^{c\dagger}_S(z\wedge (y+ {\cal K}))$ shows that the time
evolution is a
${\cal G}_S$  gauge transformation. Thus the gauge classes of  ${\cal
K} ^0$
and ${\cal Z}^1$ are $t$-independent. The   action  then can be
written as
$$S[{\cal K} ^0,{\cal Z}^1]={T\over 2 g^2} \int _{} {\cal
K} ^0\wedge \dot{{\cal Z}^1} d\sigma + ({\cal K}^0+d^{c\dagger} {\cal
Z}^1d\sigma+y)^2\wedge w+ ({\cal K}^0+d^{c\dagger} {\cal
Z}^1d\sigma+y)\wedge
y\wedge w~,$$
where $T$ is the length of  ${\bf T}^1_t$ and  integration runs over
$S\times
{\bf
T}^1_\sigma$.  We see that
$t$-dependence of
gauge invariant observables is trivial.  Thus one can expect that
%\localiz\
  is true  independently  of  the
assumption $i(\partial _t)x=0$.

The localization formula \localiz\ is specific for the factor
structure of the
target space
$M=S \times T^2$. For such  target spaces the interaction can be
removed by choosing an appropriate gauge.
This situation is very similar to Chern-Simons.

\newsec{Relation to $N=2$ topological strings}

\subsec{AKS theory and complexified K\"ahler cone}

There is a crucial difference between AKS theory and $N=2$
topological strings.
As described, AKS theory is defined on the {\it real} K\"ahler cone.
Similarly,
it is natural to think of $x$ and $K$ as real. On the other hand, the
$N=2$
TCFT is naturally
defined on the {\it complexified}  K\"ahler cone, where to  a real
positive
K\"ahler form one adds an imaginary antisymmetric tensor $B_{i\bar
j}$  ( the
$\theta$-term ).

As we just  have seen, AKS  is  background independent.  Essentially,
this is
a consequence of   flatness of the bundle of  massless states.  The
flat
connection $s(\partial _a(\omega ), \partial _b(\omega ))$ naturally
appears in
the transformation \newfield\ ~which leaves the action invariant.
Equally
important is that $s(\partial _a(\omega ), \partial _b(\omega ))$
preserves the
constraint $d^{c\dagger}x=0$.  A characteristic feature of  this
connection is
that  the covariantly constant  vector fields
$x(\omega )$
scale as\foot{In general, the covariantly constant sections of
$H^{p,q}$-bundle scale as $\lambda ^{p+q}$ as $\omega \rightarrow
\lambda
\omega$.} $x(\lambda \omega )\rightarrow \lambda^2 x(\omega )$
with respect to it.
In turn,
this\foot{To be rigorous, one has to refer here to the similar
statement about
the massive field $K'$ (see \backind ).} leads to the parabolic law
\strcoco\
for the volume-dependent string coupling constant $g(V)$.

\subsec{Classical action}

In general there is no localization formula. The   (Lagrangian)  path
integral
for AKS is given by
\eqn\pathint{e^{-\Gamma[x; \omega]}=e^{-S_0[x; \omega]}\int {\cal D}K
e^{-{\cal
A}[K,x;\omega]}~~~~,}
where we extracted the classical action $S_0[x; \omega]$ and ${\cal
A}[K,x;\omega]$
is given by \backind.
$\Gamma[x; \omega]$ is nothing else but an effective action for SQM
(or semiclassical limit for TCFT).  Below we
will prove this statement at the tree level.  Also, we will explain
how the
global background independence of AKS is translated into the global
background  independence of  SQM.

It is instructive to compare the perturbation theory for TCFT coupled
to
gravity and
AKS. The latter is formulated in terms of
Yukawa couplings $C_{abc}$ and propagators
$\Delta_{nm}$, $\Delta_n$ and  $\Delta$  (for the discussion on
perturbation
theory see \bcov\ \foot{In \bcov\
the propagators are denoted as $S_{nm}$, $S_n$ and  $S$.}) .
It was suggested in \bcov\ that introducing a {\it dilaton} field $y$
the
tadpole $\Delta_n$ and  ``blob'' $\Delta$ can be interpreted as
ordinary propagators $\Delta_n=\Delta_{ny}$ and $\Delta=\Delta_{yy}$
{}.

Below we will see how
the analog of  these operations appear in the perturbation theory for
AKS.

On one hand the AKS action evaluated
on the classical trajectory is written in terms of {\it massive}
propagator
$\Pi (\cdot)$.  On the other hand, the SQM amplitudes should be
expressed entirely in terms of string propagator $D(\cdot)$, defined
as
\eqn\massless{D(x)=\Lambda(x)-\Pi(x)}
It is clear that $D$ is well defined on cohomology, namely for $x$
being
$d$-closed
$D(x)$ is also $d$-closed, while for $x$ being $d$-exact
$D(x)$ is also $d$-exact.

Let us compute the classical action  in perturbation series
$S=S^{(3)} +S^{(4)}
+S^{(5)} + ...$
and rewrite it in terms of $D(\cdot)$.
The solution of
the classical equation of motion \aks\
is given by \Sol. Plugging this expression into the action we obtain
\eqn\classact{ S_0 [x;\omega]={1 \over g_{\omega}^2 }\left(  {1 \over
6}\int
x^3 - {1 \over 8} \int x^2 \Pi (x^2) +
{1 \over 8} \int x^2 \Pi(x\Pi(x^2)) + ... \right) ~.}
The classical action is given as series in terms of $massive$
propagator $\Pi$.
Taking into account the relation \massless\
one can rewrite the second term as
\eqn\four{S^{(4)}= {1 \over 8} \int x^2 \Pi (x^2)=
 {1 \over 4!} \left(  3 \int x^2 D(x \wedge x) -4 (\Lambda x) \int
x^3 \right)
}
It is important that $\Lambda x$ is a {\it number} and therefore it
can be
taken outside the integral.
This expression should be compared with perturbation series of $N=2$
TCFT
\bcov\ which is given by
\eqn\fourtop{S^{(4)}_{N=2}={x^a x^b x^c x^d \over 4!} \left(
 \sum _{3} C_{ab}^n \Delta_{nm} C_{cd} ^m -   \sum_{4} \partial_a K
C_{bcd}
\right)}
Comparing these two equations we can indeed identify term by term.
$C_{abc}$ are Yukawa couplings, while $\Delta_{nm}$ are matrix
elements of
propagator
$\Delta(\cdot)$.
The relation $\partial_a K=\Lambda x_a$ follows the fact that
K\"ahler potential is given by $-{\rm Log}({\rm Vol}_{\omega})$.

There are some new features which appear at the next order in the
perturbation series.
Namely, there is a diagram whose contribution can be interpreted as
coming from
a tadpole $D_n$.
Indeed the fifth order contribution to the action is
\eqn\fif{\eqalign{
S^{(5)}= &{1 \over 5 !} \left( 15 \int (D(x \wedge x)) ^2 x - 30
(\Lambda x)
\int x^2 D(x \wedge x)  \right) + \cr
&{1 \over 5 !} \left(20  (\Lambda x)^2 \int x^3  +10 [{1 \over
2}\Lambda^2
(x^2)- \Lambda D (x^2) ] \int x^3   \right) \cr } }
In deriving \fif\ it was important that ${1 \over 2}\Lambda^2 (x^2)-
\Lambda D
(x^2)$ is a  {\it number} and therefore
it can be taken outside the integral.  Again, this expression should
be
compared with perturbation series of $N=2$
TCFT
\eqn\fiftop{\eqalign{S^{(5)}_{N=2}=&{x^a x^b x^c x^d x^e \over 5!}
(\sum_{15} C_{ab}^n \Delta_{nm} C^{mk}_c \Delta_{kp} C_{de}^p -
 2 \sum_{5} \partial_e K \sum_{3}   C_{ab}^n \Delta_{nm} C_{cd} ^m +
\cr
& 2 \sum_{10} \partial_a K \partial_b K  C_{cde} +
\sum_{10}  ( \Delta_n - \partial_m K \Delta_n ^m ) C^n _{ab}
C_{cde}  \cr} }
Again, one can identify term by term
\eqn\tad{{1 \over 2}\Lambda^2 (x^2)- \Lambda D (x^2) =  ( \Delta_n -
\partial_m
K \Delta_n ^m ) C^n _{ab}  x^a x^b  .}
The existence of tadpole $\Delta_n$ and dilaton-dilaton propagator
$\Delta$ are
related to the possibility
of constructing the numbers out of $\Lambda(\cdot)$, $D(\cdot)$ and
$x$.
In fact there are only three irreducible possibilities (for a
$3$-fold)
$~\Lambda (x) ,~
  -{1 \over 2}\Lambda^2 (x^2)+ \Lambda D (x^2) $ and
  $ {2 \over 3}\Lambda^3 (x^3)- \Lambda^2 D (x^3) $.
The dilaton-dilaton propagator $\Delta$ is related to the last
combination
(the explicit expression is quite complicated and we won't present it
here).
These calculations suggest that at every order in perturbation theory
the classical action is expressible entirely in terms of massless
propagators and Yukawa couplings.

The perturbation theory for AKS is identical to perturbation theory
of $N=2$
TCFT
at least at the tree level. It is tempting to suggest that this
similarity
persist at the loop level and one may just borrow the perturbation
series for
$N=2$
TCFT in order to construct loop corrections to AKS theory.
For example the one loop correction to the one point correlation
function
should be given as follows
$\langle x \rangle_1 =x^a C_{abc} \Delta^{bc}$. We do not know how to
prove
this suggestion.

There seems to be a contradiction.
The interpretation of of AKS as SQM seems to be at odds
with the appearance of the dilaton field in the perturbation theory.
There is nothing like a dilaton field  in SQM.
We can suggest the following resolution of this puzzle.
Let us introduce an $x$-dependent factor
\eqn\xvolume{F(x)=1+ \Lambda(x) +{1 \over 2 } [-{1 \over 2}\Lambda^2
(x^2)+
\Lambda D (x^2)]+ ...~~.}
Using this factor one can rearrange the perturbation series for the
classical
action as\foot{The similar statement is valid in
topological string theory.  The function $F(x)$ is given as follows
$F(x)=1+x^i \partial_i +{1 \over 2} x^a x^b  C^n _{ab} ( \Delta_n -
\partial_m
K \Delta_n ^m )+...$}
\eqn\newclass{\eqalign{S_0[x;\omega]=& F(x)^2
\Big\{  {1 \over 3!} \left({1\over F(x)} \right)^3 \int x^3 +\cr
{3 \over 4!} \left({1\over F(x)}\right)^4\int x^2& D(x^2)+
{15 \over 5!} \left({1\over F(x)}\right)^5\int x (D(x^2))^2 +...
\Big\} \cr }}
Now it is clear that one can make {\it field}-dependent
renormalization of the
external legs
and
the coupling constant
\eqn\ren{x \longrightarrow \phi={x \over F(x)}~~{\rm and}~~~g^2
\longrightarrow
{g^2 \over F(x)^2}}
and recover the conventional perturbation series for SQM.

\subsec{Yukawa couplings}

We are going to prove that classical action $S_0
[x;~\omega]$ is in fact a generating functions for string amplitudes.
It is convenient to write a relation not for the classical action but
for
perturbed Yukawa coupling.
\eqn\genfun{\partial_{x^i} \partial_{x^j} \partial_{x^k}  {1 \over
g^2
_{\omega}} S_0 [x;~\omega]=C_{ijk}[x;~ \omega]=
\sum_{N \geq 0}{1 \over N!}x^{p_1}...x^{p_N} A_{ijkp_1...p_N}
[0;~\omega]}
Let us show that $A_{ijkp_1...p_N} [0;~\omega]$ are $(N+3)$ point
correlation
functions, given as
\eqn\flatyuk{A_{ijkp_1...p_N}
[0;~\omega]=(-1)^N\partial_{p_1}...\partial_{p_N}
C_{ijk}}
 where the derivatives $\partial_{p_i}$ are with respect to the {\it
flat
coordinates} $(\tau ^1,\ldots,\tau ^n)$ of the point $\omega$ on the
K\"ahler
cone K.  This is equivalent to

\vskip3mm
\eqnn\bincor
\hskip 5pc \Boxit{\vbox{\hsize 14pc  $$C_{ijk}[x;~
\omega]=C_{ijk}(x^1-\tau
^1,\ldots,x^n-\tau ^n)$$}}\hskip5pc \raise1pc \hbox{\bincor}

Following the discussion in \bcov\ one can represent the perturbed
Yukawa
couplings as
\eqn\yuk{C_{ijk}[x;~ \omega]=\partial_{x^i} \partial_{x^j}
\partial_{x^k}  {1
\over g^2 _{\omega}} S_0 [x;~\omega]=   {1 \over g^2 _{\omega}}
     \int  \partial_{x^i} K_0
\partial_{x^j} K_0
\partial_{x^k} K_0~.}

Let us consider the solution $x(t)$ of  the equation \eoc\ such that
$x(\tau
)=x$ at the point $\omega=(\tau ^1,\ldots,\tau ^n)$.
As it was discussed in Section 4, for the   flat coordinates
$x(t)=x^i
(t)e_i=(t^i - t^i_0)e_i\, ,$ where $e_i$ are the coordinate vectors
(cf.
\linsec\ ). Obviously, the parameters $t_0^i$ are related to $x^i,\,
\tau ^i$
by
\eqn\xttau{x^i=\tau ^i-t_0^i.}

Consider the Yukawa coupling evaluated on the solution $x (t)$. Under
the small
variation of  K\"ahler structure $K_0$ transforms according to
\newfield b\
and  therefore
\eqn\derivative{\partial_{x^i} K_0 \rightarrow  \partial_{x^i} \tilde
K_0 =
\partial_{x^i} K_0- m(\delta \omega,\partial_{x^i} K_0 )}
It  is easy to see that $C_{ijk}[x(t);~ \omega (t)]$ does not depend
on $t$.
Indeed,
\eqn\three{\eqalign{C_{ijk}[x;~ \omega]=&{1 \over g^2_{\omega}}
                                 \int  \left( \partial_{x^i} \tilde
K_0
\partial_{x^j}  \tilde K_0 \partial_{x^k} \tilde K_0  + \sum_{\rm
perm}
m(\delta \omega, \partial_{x^i} \tilde K_0)   \partial_{x^j}  \tilde
K_0
\partial_{x^k} \tilde K_0 \right)= \cr
&= {1 + 2 \Lambda (\delta \omega)\over g^2 _{\omega}}
                                 \int   \partial_{x^i} \tilde K_0
\partial_{x^j} \tilde K_0 \partial_{x^k} \tilde K_0   =
C_{ijk}[\tilde x;~
\tilde \omega].
}}
(It is important that $C_{ijk}[x;~ \omega]$ is written in flat
coordinates.
Otherwise the additional Jacobian factors  would appear in \three.)

Now we can apply the argument  used already in Section 3 (and
explained in
detail in Section 4).  There is a point  $\tilde \omega
=(t_0^1,\ldots ,t_0^n)$
on the K\"ahler cone K  such that $x^i(\tilde \omega )=0$. At this
point,
obviously, $\partial_{x^i}  K_0[x(t_0)]=\partial_{x^i} x(t_0)=e_i$.
Therefore
\eqn\intin{C_{ijk}[0;~\tilde \omega]={1 \over g^2 _{\tilde
\omega}}\int
e_i \wedge e_j \wedge e_k}
can be interpreted in terms of  intersections in $H^{1,1}(M)$.
Moreover, \three\ together with \xttau\ show that the {\it perturbed}
Yukawa
coupling $C_{ijk}[x; ~\omega]$ computed for the K\"ahler structure
$\omega
=(\tau ^1,\ldots,\tau ^n)$
coincides with Yukawa coupling  $C_{ijk}[0;~\tilde \omega]$ for the
K\"ahler structure $\tilde \omega=(\tau ^1-x^1,\ldots,\tau ^n-x^n)$.
Therefore
as a function, $C_{ijk}[x; ~\omega]$ depends only on  $\tau ^i -
x^i$. This
implies \flatyuk.

\newsec{Concluding remarks}

One of the main motives of this paper is the connection on  the
Hilbert space
bundle.
It first appears it in the context of  TCFT.  The notion of
background
independence is
formulated in terms of this connection (to be precise, in terms of
{\it affine}
connection).
Background independence of  TCFT is equivalent to flatness of this
connection.
The condition of supersymmetry imposes strong constraints on the form
of the
connection which have a simple solution in the semiclassical regime.
This solution is constructed in terms of  geometric operation
$\Lambda$.
We suggest that operation $\Lambda$ can be defined for any $N=2$
field theory
(not only in the semiclassical limit) and the connection is given in
terms of
$\Lambda$. The semiclassical limit  of TCFT is SQM (details of
identification
discussed in the main body of the paper).
The states in SQM are identified with harmonic forms and
therefore  the connection in question is the connection on the leaves
of Hodge
foliation. This connection turns out to be flat.

The same connection appears in AKS. It allows
one to relate theories for  different K\"ahler structures.
It enters into the formulation of background independence.
AKS is a gauge theory with the gauge group ${\cal SD}iff$.
The gauge symmetry is free of anomalies which can be checked
by direct computations.
Unfortunately, we were not able to construct the gauge invariant
observables which may be a good direction to pursue.

It is natural to compare AKS with Chern-Simons theory.
Chern -Simons is a topological theory and its hamiltonian is equal to
zero.
AKS is K\"ahler topological theory and its hamiltonian
differs from zero. This hamiltonian determines a unique
dynamics on the phase space. In case when the target space has factor
structure
$M={\bf T}^2 \times S$, AKS reduces to a free theory with
constraints.
In this case AKS can be quantized and one can derive a simple
localization
formula.

\bigskip
\noindent
{\bf Acknowledgements}.

We would like to thank A. Astashkevitch, A. Alekseev,
S. Axelrod, D. Kazhdan, A. Losev, C. Vafa and S.T. Yau for valuable
discussions.
This research was supported in part by NSF grant
PHY-92-18167, NSF 1994 NYI award,  DOE 1994 OJI award and a
Packard Foundation.

\appendix{A}{ Differential Geometry of $m(A,B)$}
\subsec{ Hodge identities and $sl(2)$}

Let us  introduce the operator   $L$ of multiplication by K\"ahler
form $\omega
_{i\bar j}$, the operator $\Lambda$ of contraction with bivector
$\omega
^{i\bar j}$ and the operator $J$ which acts on $(p,q)$ forms as $dim
M-(p+q)$.
The operators $\Lambda$, $L$  and
$J$ form the $sl(2)$  algebra:
\eqn\slI{\eqalign{
[\Lambda ,L]&=J \cr
[J,L]&=2L\cr
[J,\Lambda ]&=-2\Lambda \cr
}}

 The commutation relations between $L$, $\Lambda$
and $\partial$, $\bar \partial$, $\partial^{ \dagger}$, $\bar
\partial^{
\dagger}$ are known as the Hodge identities
\eqn\hodgeid{\eqalign{&[\partial, \Lambda]=-\bar \partial^{ \dagger}
{}~~{\rm
and}~~~[\partial ^{\dagger}, L]=\bar \partial\cr
&[\bar \partial, \Lambda]=\partial^{ \dagger} ~~{\rm and}~~~[\bar
\partial^{
\dagger}, L]=-\partial \cr }}
(see  \grhar).  These relations imply the Hodge identities for the
Laplacians:
\eqn\HoLap{2\Delta _{\partial}=2\Delta _{\bar \partial}=\Delta _{d}.}
Also,
\hodgeid\ means that $sl(2)$ defined by \slI\  preserves ${\rm
Ker}\Delta$,
i.~e.~sends  harmonic forms to harmonic ones.  Thus the space of all
harmonic
forms has a decomposition (Lefshetz decomposition) as a direct sum of
irreducible representations of this algebra.  The highest weight
vectors
annihilated by $\Lambda$ are called primitive forms.  The following
formula is
useful in applications.
If $p$ is a primitive form of rank $r$, then (~$\omega $ is the
K\"ahler
form~)
\eqn\lapow{
\Lambda  (\omega ^kp)=k(n-k-r+1)\omega ^{k-1}p.
}

\subsec{Properties of  $m(A,B)$}
A bilinear symmetric operation on differential forms  $m(A,B)$ is
defined
as follows
\eqn\mab{
m(A,B)= \Lambda (A\wedge  B)-(\Lambda A)\wedge B-A\wedge \Lambda B
}
This operation has several remarkable properties, summarized below.

 Since $\Lambda$ makes contraction of {\it two} indices, there is a
formula
\eqn\na{\eqalign{
\Lambda (ABC)=&
\Lambda(AB)C+\Lambda(CA)B+\Lambda(BC)A-\cr
& \Lambda(A)BC-A\Lambda(B)C-AB
\Lambda(C)~,
\cr }}
which is equivalent to
\eqn\difproD{
m(AB,C)=Am(B,C)+m(A,C)B.
}
Thus the operation $m(\cdot ,\cdot )$ {\it differentiates} the
multiplication
of forms.

It is convenient  for us to introduce four differential operators
\eqn\fodio{\eqalign{
d=\partial + \bar \partial ~~~~~&{\rm and }~~~~~d^c=\partial - \bar
\partial~,\cr
d^{\dagger}=\partial ^{\dagger}+ \bar \partial ^{\dagger}~~~~~&{\rm
and
}~~~~~d^{c\dagger}=\partial ^{\dagger}- \bar \partial
^{\dagger}~\cr}}
instead of  $\partial ,\,\bar \partial  ,\, \partial ^{\dagger}$ and
$ \bar
\partial ^{\dagger}$.
Acting on $m(A,B)$ by $d$ one gets
\eqn\me{\eqalign{
d m(A,B)&=d^{c \dagger}(A \wedge B)-(d^{c \dagger}A)\wedge B-A\wedge
d^{c
\dagger}B+ m(dA,B) + m(A,dB)\cr
}}
Suppose now that $K$ is a {\it harmonic} form; then \me\ means that
\eqn\crur{\eqalign{
[d^{c \dagger},K]B&=dm(K,B)-m(K,dB) \cr
[d^{ \dagger},K]B&=d^c m(K,B)-m(K,d^c B) \cr
}}
If we deform the K\"ahler form $\omega$ by adding $K$ to it,
$\Lambda$ changes.
To the first order,
\eqn\varl{
\delta \Lambda ={1\over 2}[\Lambda ,[\Lambda , K]].}
Then one has a relation, ``dual" to \crur:
\eqn\crurII{\eqalign{
[d,\delta \Lambda ]B&=d^{c\dagger}m(K,B)-m(K,d^{c\dagger}B)\cr
[d^c,\delta \Lambda ]B&=d^{\dagger}m(K,B)-m(K,d^{\dagger}B)\cr
}}
We will prove here only the first relation:
\eqn\lhsidid{\eqalign{
[d,\delta \Lambda ] B&={1\over 2}[d,[\Lambda ,[\Lambda ,
K]]]B={1\over
2}d\bigl(\Lambda ^2(KB)-2\Lambda (K\Lambda B)  +K\Lambda ^2B\bigr)
\cr
&-{1\over 2}\bigl(\Lambda ^2(KdB)-2\Lambda (K\Lambda dB)  +K\Lambda
^2dB\bigr)
\cr
&={1\over 2}\bigl(2\Lambda d^{c\dagger}(KB)-2d^{c\dagger}(K\Lambda B)
-2\Lambda
(Kd^{c\dagger}B) +2K\Lambda d^{c\dagger}B\bigr)
}}
On the other hand,
\eqn\rhsidid{\eqalign{&
d^{c\dagger}\bigl (\Lambda (KB)-(\Lambda K) B-K(\Lambda
B)\bigr)=\cr
&\Lambda
d^{c\dagger}(KB)-d^{c\dagger}(K(\Lambda B))-[d^{c\dagger}, \Lambda
K]B +
(\Lambda K) d^{c\dagger}B \cr
&=\Lambda d^{c\dagger}(KB)-d^{c\dagger}(K(\Lambda B))+ (\Lambda K)
d^{c\dagger}B
}}
(We used that $K$ is the harmonic 2-form,  so $[d^{c\dagger}, \Lambda
K]=0$.)

\appendix{B}{A space of massless modes ${\cal H}$}
By definition, the massless modes $x$ satisfy  two equations:
\eqn\mamo{
dx=0~~~~~~{\rm and}~~~~~~~d^{c\dagger}x=0}
The space of solutions of  \mamo\ we call  ${\cal H}$. It is defined
for every
K\"ahler form $\omega$ and depends on it. Since the differential
operators $d$
and  $d^{c\dagger}$ commute with each other the space ${\cal H}$ has
to be
infinite-dimensional: it contains a subspace ${\cal T}$ which
consists of forms
$dd^{c\dagger}t$.

\noindent
{\it A quotient ${\cal H}/ {\cal T}$ is finite
dimensional. For a
given complex structure $J$ it can  be identified with harmonic forms
canonically.}

\noindent
Indeed, given a complex structure, one has a  Hodge
decomposition of  a
$d$-closed form $x$:
\eqn\Hod{x=h_x+d\beta ,}
where $h_x$ is harmonic and $d^\dagger \beta=0$. The second equation
in \mamo\
tells us that $dd^{c\dagger}\beta =0$, so $d^{c\dagger}\beta
=h_{d^{c\dagger}\beta}+d\Gamma$,  where  $h_{d^{c\dagger}\beta}$ is
harmonic.
Let us prove that  this harmonic piece  is always zero. Indeed,  for
any
harmonic form $\omega$ we have
$$\int \omega \wedge h_{d^{c\dagger}\beta}=\int \omega \wedge
d^{c\dagger}\beta
=0.$$
Thus  $d^{c\dagger}\beta =d\Gamma$. To prove that $d\Gamma =0$ we act
by
$d^{\dagger}$ on both sides of this formula to obtain
$d^{\dagger}d\Gamma =0
\Longrightarrow d\Gamma =0.$ Indeed, $$0=\langle \Gamma ,
d^{\dagger}d\Gamma
\rangle =\langle d\Gamma , d\Gamma \rangle \geq 0 \Longrightarrow
d\Gamma =0.$$

So always $d^{c\dagger}\beta =0$ and $d^{\dagger}\beta =0$, hence  $
d*\beta
=d^{c}*\beta =0$. Now we can write $*\beta =h_{*\beta}+d\chi$ where
$*\beta$ is
harmonic.  The harmonic component  can be disregarded, in fact, since
$\beta$
itself  is defined modulo harmonic forms. So finally $*\beta =d\chi$;
then it
follows from the $dd^c$ -lemma  that
$*\beta =dd^c \gamma$ and  $\beta =d^{c\dagger}d^{\dagger}*\gamma$.
The
decomposition \Hod\ gives  $x=h_x+ dd^{c\dagger}t_x $ and the lemma
is proved.

\appendix{C}{Hodge foliation on the space of K\"ahler  forms.}
Let us consider  the infinite-dimensional  linear space ${\cal
K}^{1,1}(M)$ of
K\"ahler forms on $M$. There is a natural distribution  on  ${\cal
K}^{1,1}(M)$
with a fiber at point $\omega \in {\cal K}^{1,1}(M)$ defined as a
space
H$_\omega$ of harmonic forms with respect to $\omega$.
Here we want to prove that this distribution is integrable ---
i.~e.~produces a
foliation of  ${\cal K}^{1,1}(M)$ by  finite-dimensional leaves. Each
leaf is,
loosely speaking, a lift  to ${\cal K}^{1,1}(M)$ of the cohomological
K\"ahler
cone K.

In particular, this  fact implies that one can introduce the
coordinates on
${\cal K}^{1,1}(M)$ such that  $n={\rm dim}H^{1,1}(M)$ coordinate
lines always
have the harmonic tangent vectors $\partial _a$   (we use the
canonical
identification ${\cal K}^{1,1}(M)\cong T{\cal K}^{1,1}(M)$).

To prove the   statement  we check  the Fr\"obenius integrability
condition
for a pair of vector fields $\xi$ and $\eta$ such that $\xi (\omega
)$ and
$\eta (\omega )$ are harmonic forms for all $\omega \in {\cal
K}^{1,1}(M)$.
Let us introduce (~global~) coordinates on  ${\cal K}^{1,1}(M)$ using
the Hodge
decomposition with respect to some particular $\omega _0$. It
suffices to show
that the commutator $[\xi ,\eta ](\omega _0)$ is harmonic with
respect to
$\omega _0$.  For the infinitesimal variation $\omega =\omega
_0+\delta \omega$
one can write:
\eqn\vahaf{\xi (\omega ) =\xi (\omega _0)-s(\delta \omega ,\xi(\omega
_0))+h,}
where $h$ is harmonic w.~r.~t.~$\omega _0$ and the operation
$s(\cdot,\cdot)$
is defined as  $s(A,B)=P_{{\rm Ker} (d)} m(A,B)=m(A,B) - \Pi(AB)$
(cf.
\decomp). The relation \vahaf\  follows from the properties of
$m(A,B)$
discussed above.  Indeed,  $\xi (\omega)$ is $d$-closed
($\Longleftarrow$
\crur ) and annihilated by $d^{\dagger}$ ($\Longleftarrow$ \crurII)
modulo
$o(\delta \omega ^2)$ terms. Therefore, the commutator  $[\xi ,\eta
]$ equals
$$[\xi ,\eta ](\omega _0)= s(\xi(\omega _0),\eta (\omega _0))-s(\eta
(\omega
_0) ,\xi(\omega _0))+h_{\eta ,\xi}=h_{\eta ,\xi}$$
i.~e. it is harmonic.

In the tangent bundle ${\cal TL}$ to the leaf  ${\cal L}$ one can
introduce a
connection
\eqn\connh{D_\xi=\partial _\xi -s(\xi ,\cdot)~.}
Arguments similar to ones in the Section 4.2 show  that  $D_\xi$ is
flat.  The
connection $D_\xi$ is not a metric one. The connection compatible
with the
Riemann metrics
$$ \langle \eta \mid \xi \rangle={1\over {\rm Vol}_\omega} \int  (*
\eta \wedge
\xi) $$
on (real) forms is given by $~~\partial _\xi -{1\over 2}s(\xi
,\cdot)$.
Unfortunately,  this is not a connection on  ${\cal TL}$ since the
parallel
transport by it does not send harmonic forms to harmonic ones.  Then,
though
the metrics connection descends to {\it cohomology}  as a connection
in the
tangent bundle to the cohomological K\"ahler cone, it is not flat.

\listrefs

\end

Finally, we want to obtain the localization formula for  any $M$, not
necessarily of the type $S\times {\bf T}^2$.  For that purpose it is
natural
to use the Lagrangean path integral (based on the BV formulation
studied in
Section 3.4). We hope that  the above discussion was convincing
enough to show
that AKS is a {\it bona fide} quantum field theory and that the path
integral
is well defined.

The key feature now is the existence of global fermionic (BRST)
symmetry
$\delta _{BRST}$. On general grounds we conclude that the path
integral can be
localized to the fixed points set of $\delta _{BRST}$ which coincides
with the
set of solutions of  classical equations of motion.  In turn, these
solutions
are parameterized by the parameter $\alpha$ of the gauge group ${\cal
G}_M$.
Thus the path integral is reduced to integration over  ${\cal G}_M$.

To implement this idea, it is convenient to use the ``background
independent''
action $\cal A$ which appeared in Section 3.6. So we take a
decomposition
${\cal K}={\cal K}_0+{\cal K}'$ where ${\cal K}_0$ solves the
equations of
motion and $ {\cal K}'=d^{c\dagger}{\cal Z}'$ is a transversal mode
({\it not}
annihilated by the operator  $L=dd^{c\dagger}+d^{c\dagger}({\cal
K}_0\wedge
d^{c\dagger})~$). The action is
$$S[ x,{\cal K};\,  \omega ]= S_0[ x ;\,  \omega]+{\cal A} [{\cal
K}', x ;~
\omega]  =
   S_0[ x ;\,  \omega ]+ {1 \over g^2}
   \int \Bigl({1 \over 2} {\cal K}' {1  \over
d^{c\dagger}} D {\cal K}'   +  {1 \over 6} {\cal K}' \wedge{\cal K}'
\wedge{\cal K}' \Bigr),$$
where $ S_0[ x ;\,  \omega ]$ is the action $S[{\cal K}]$ computed on
the
clasical solution  $ {\cal K}_0$ and  $D=d+[d^{c\dagger},\,{\cal
K}_0]$.
Assuming, that BRST invariance remains at the quantum
level we can change the action by BRST trivial
piece.
Let us apply the Duistermaat-Heckman \ref\duheck{Some reference}
localization principle to the
Lagrangean path
integral of the BRST-invariant observable ${\cal O}[{\cal K}]$  ($
{\cal Z}'$
runs over transversal modes and $ {\cal K}_0$ runs over solutions)
\eqn\Lpi{\eqalign{
\int &{\cal DK}_0\,{\cal D Z}'\,{\cal O}[{\cal K}]\,e^{\, -S[{\cal
K}]}=e^{\,-S\big[{\cal K}_0[x]\big]}\int  {\cal DK}_0\,{\cal
DZ}'\,{\cal
O}[{\cal K}]\,e^{\,-{\cal A} [{\cal Z}', x ]}\cr
&=e^{\,-S\big[{\cal K}_0[x]\big]}\int {\cal DK}_0\,{\cal DZ}'\,{\cal
O}[{\cal
K}]\,\delta [Dd^{c\dagger}{\cal Z}']\,[{\rm Ber}\,'d^{c\dagger}D]\cr
&=e^{\,-S\big[{\cal K}_0[x]\big]}\int  {\cal D}\alpha \,{\cal
O}\big[{\cal
K}_0[x]\big]\cr
}}